\documentclass[10pt,onecolumn]{IEEEtran}

\usepackage{makeidx}
\usepackage{amsmath}
\usepackage{epsfig}
\usepackage{amssymb}
\usepackage{textcomp}
\usepackage{pstricks,pst-text}
\usepackage{algorithm}
\usepackage{algorithmic}
\usepackage{datetime}
\usepackage{setspace}
\usepackage{stfloats}
\usepackage{subcaption}
\usepackage{cases}
\usepackage{cite}
\usepackage{watermark}

\newcommand{\mbf}{\mathbf}

\newcommand{\bzero}{\mbf{0}}

\def\b1{{\mathbf 1}}

\newcommand{\balpha}{\boldsymbol\alpha}
\newcommand{\bbeta}{\boldsymbol\beta}
\newcommand{\bomega}{\boldsymbol\omega}
\newcommand{\bphi}{\boldsymbol\phi}

\def\bgamma{{\mbox{\boldmath{$\gamma$}}}}
\def\bGamma{{\mbox{\boldmath{$\Gamma$}}}}
\def\bLambda{{\mbox{\boldmath{$\Lambda$}}}}
\def\blambda{{\mbox{\boldmath{$\lambda$}}}}

\def\bphi{{\mbox{\boldmath{$\phi$}}}}

\def\bSigma{{\mbox{\boldmath{$\Sigma$}}}}

\newcommand{\cA}{\ensuremath{\mathcal{A}}}
\newcommand{\cB}{\ensuremath{\mathcal{B}}}
\newcommand{\cC}{\ensuremath{\mathcal{C}}}

\newcommand{\cG}{\ensuremath{\mathcal{G}}}

\newcommand{\cN}{\ensuremath{\mathcal{N}}}
\newcommand{\cO}{\ensuremath{\mathcal{O}}}

\newcommand{\cR}{\ensuremath{\mathcal{R}}}
\newcommand{\cS}{\ensuremath{\mathcal{S}}}

\newcommand{\diag}{\mbox{\rm diag}}

\def\eg{{e.g.,\ }}
\def\ie{{i.e.,\ }}

\newcommand{\betab}{\begin{tabbing}}
\newcommand{\entab}{\end{tabbing}}
\newcommand{\beitem}{\begin{itemize}}
\newcommand{\enitem}{\end{itemize}}
\newcommand{\bea}{\begin{array}}
\newcommand{\ena}{\end{array}}
\newcommand{\beq}{\begin{equation}}
\newcommand{\enq}{\end{equation}}
\newcommand{\beqa}{\begin{eqnarray}}
\newcommand{\enqa}{\end{eqnarray}}
\newcommand{\beqan}{\begin{eqnarray*}}
\newcommand{\enqan}{\end{eqnarray*}}
\newcommand{\beenum}{\begin{enumerate}}
\newcommand{\enenum}{\end{enumerate}}
\newcommand{\DL}{\begin{dashlist}}
\newcommand{\DLE}{\end{dashlist}}

\newcommand{\ba}{{\ensuremath{\mathbf{a}}}}
\newcommand{\bb}{{\ensuremath{\mathbf{b}}}}
\newcommand{\bc}{{\ensuremath{\mathbf{c}}}}
\newcommand{\bd}{{\ensuremath{\mathbf{d}}}}
\newcommand{\be}{{\ensuremath{\mathbf{e}}}}

\newcommand{\br}{{\ensuremath{\mathbf{r}}}}

\newcommand{\bt}{{\ensuremath{\mathbf{t}}}}

\newcommand{\bx}{{\ensuremath{\mathbf{x}}}}

\newcommand{\bz}{{\ensuremath{\mathbf{z}}}}

\newcommand{\bA}{{\ensuremath{\mathbf{A}}}}
\newcommand{\bB}{{\ensuremath{\mathbf{B}}}}
\newcommand{\bC}{{\ensuremath{\mathbf{C}}}}

\newcommand{\bE}{{\ensuremath{\mathbf{E}}}}
\newcommand{\bF}{{\ensuremath{\mathbf{F}}}}
\newcommand{\bG}{{\ensuremath{\mathbf{G}}}}
\newcommand{\bH}{{\ensuremath{\mathbf{H}}}}
\newcommand{\bI}{{\ensuremath{\mathbf{I}}}}
\newcommand{\bJ}{{\ensuremath{\mathbf{J}}}}

\newcommand{\bP}{{\ensuremath{\mathbf{P}}}}

\newcommand{\bR}{{\ensuremath{\mathbf{R}}}}

\newcommand{\bT}{{\ensuremath{\mathbf{T}}}}
\newcommand{\bU}{{\ensuremath{\mathbf{U}}}}
\newcommand{\bV}{{\ensuremath{\mathbf{V}}}}

\newcommand{\bX}{{\ensuremath{\mathbf{X}}}}

\def\bSigma{{\mbox{\boldmath{$\Sigma$}}}}
\def\bSum{{\mbox{\boldmath{$\sum$}}}}
\def\cSigma{\emph{\mbox{\boldmath{$\Sigma$}}}}

\def\btheta{{\mbox{\boldmath{$\theta$}}}}

\def\bEta{{\mbox{\boldmath{$\eta$}}}}

\def\wrt{{w.r.t.\ }}

\newcommand{\Cramer}{Cram\'{e}r}

\newcommand{\barN}{{\ensuremath{\bar{N}}}}

\newcommand{\firstAuthor}       {Raj~Thilak~Rajan*,~\IEEEmembership{Student Member,~IEEE}}
\newcommand{\secondAuthor}      {Alle-Jan~van~der~Veen,~\IEEEmembership{Fellow,~IEEE}}

\newcommand{\theTitle}	{Joint ranging and synchronization \\ for an anchorless network of mobile nodes}
\newcommand{\ieee}      {\emph{IEEE}}
\title  {\theTitle}
\author {
\firstAuthor\ \ and \ \secondAuthor
\thanks{R.T.Rajan is with Netherlands Institute for Radio Astronomy (ASTRON), Dwingeloo, The Netherlands (email: rajan@astron.nl) and TU Delft, Delft, The Netherlands}
\thanks{A.-J. van der Veen is with TU Delft, Delft, The Netherlands (email: a.j.vanderveen@tudelft.nl)}
\thanks{This research was funded in part by the STW OLFAR project (Contract Number: 10556) within the ASSYS perspectief program.}
\thanks{A part of this work has been published in \ieee\ CAMSAP 2011 \cite{rajanCAMSAP11} , \ieee\ ICASSP 2012 \cite{rajanICASSP12} and EUSIPCO 2013 \cite{rajanEUSIPCO13} conferences.}
}

\newcounter{remarkCounter}
\stepcounter{remarkCounter}

\newcounter{eqnCounter1}
\newcounter{eqnCounter2}
\newcounter{eqnCounter3}
\newcounter{eqnCounter4}

\newcommand{\EEPLS}     {$\text{E}^2\text{PLS}$}
\newcommand{\EEGLS}     {$\text{E}^2\text{GLS}$}

\begin{document}

\maketitle

\begin{abstract} Synchronization and localization are critical challenges for the coherent functioning of a wireless network, which are conventionally solved independently. Recently, various estimators have been proposed for pairwise synchronization between immobile nodes, based on time stamp exchanges via two-way communication. In this paper, we consider a \textit{network of mobile nodes} for which a novel joint time-range model is presented, treating both unsynchronized clocks and the pairwise distances as a polynomial function of \textit{true} time. For a set of nodes, a pairwise least squares solution is proposed for estimating the pairwise range parameters between the nodes, in addition to estimating the clock offsets and clock skews. Extending these pairwise solutions to network-wide ranging and clock synchronization, we present a central data fusion based global least squares algorithm. A unique solution is non-existent without a constraint on the cost function (\eg clock reference node). Ergo, a constrained framework is proposed and a new Constrained \Cramer\ Rao Bound (CCRB) is derived for the joint time-range model. In addition, various constraints are proposed and their effects on the proposed algorithms are studied. Simulations are conducted and the proposed algorithm is shown to approach the theoretical limits. \\

\IEEEkeywords joint estimation, position, relative position, clock synchronization, skew, offset, distance, wireless network, anchorless, motion, constrained least squares, sum constraint, nullspace constraint \end{abstract}

\section{Introduction} \label{sec:intro} The coherent functioning of wireless networks relies heavily on time synchronization among nodes \cite{sundaram05}. All nodes in a network are equipped with independent clock oscillators, which must be synchronized to a global reference, to facilitate accurate time stamping of data and synchronized communication of processed information. Clock oscillators in these nodes are inherently non-linear\cite{barnes71}, however, if calibrated astutely, can be approximated as a linear function for a small measurement time period. The unknown regression coefficients of such a model will be the clock offset and clock skew for an affine clock model. Global time synchronization within the network is then achieved by estimating all clock offsets and clock skews of the nodes and compensating the respective clocks aptly. Furthermore, when nodes are arbitrarily deployed in the field, then position estimation is often equally critical as time synchronization \cite{patwari05}. The intermediate distances between all the nodes in the network (obtained via ranging) is one of the key inputs for almost all localization techniques \eg Time Of Arrival (TOA), Time Difference of Arrival (TDOA) \cite{patwari05}, Multi-Dimensional Scaling (MDS) \cite{borg97}. When moreover the nodes are mobile, distance estimation using ranging is a challenge, particularly when the clocks of the nodes are unsynchronized.

In this article, we consider an \emph{anchorless network of unsynchronized  mobile nodes}, capable of two-way communication. All the nodes are in motion \ie \emph{mobile} during the two-way communication and hence the pairwise distances are time-varying. In addition, all the nodes are equipped with independent clocks, which are \emph{unsynchronized} \wrt some reference time \ie \emph{true} time, during the two-way communication. Finally, by the term \emph{anchorless}, we consider an autonomous and cooperative network with no external (reference) information on either time, distance or position.  Hence we assume no a priori knowledge on the nodes initial positions and/or on their respective motion. Thus, our fundamental challenge is to understand the joint variation of local time at each node and time varying pairwise distances between the cluster of nodes. After obtaining the pairwise distances at discrete intervals of time, the relative positions of the nodes at respective time instances can be obtained by applying the MDS \cite{borg97}. We assume the need for two-way communication between the nodes, but a full mesh network is not always necessary.

\subsection{Framework} We focus our attention on a two-way time stamp exchange framework, which for a fixed network of immobile nodes is a well investigated topic \cite{ieee07,Serpedin2009}. For a pair of fixed nodes capable of two-way communication with each other, the classical Two Way Ranging (TWR) model contains $2$ clock offsets, $2$ clock skews and the distance between the nodes, which results in an unsolvable five dimensional problem \cite{wu11}. However, traditionally, one clock is assumed to be the reference clock which reduces the cardinality to $3$ and given sufficient measurements, the absolute clock skew and clock offset of the second node, and its pairwise distance from the first node can be estimated. For estimating the clock errors,  maximum likelihood estimates and Low Complexity Least Square (LCLS) estimates are proposed in \cite{noh07} and \cite{leng10} respectively. A step further, joint estimation of clock parameters and the fixed distances for the entire network of nodes is proposed in \cite{rajanCAMSAP11}. However, all these propositions are based on the two-way ranging data model \cite{freris10,ieee07}, where the node positions are fixed and thus the pairwise ranges are independent of time. \emph{When the nodes are in motion, the pairwise distances are a non-linear function of time and our proposition is to approximate this continuous function as a Taylor series, for a small measurement period}. Under this context, the unknown coefficients of this monomial approximation (called range parameters) need to be estimated, which beget he pairwise distances at discrete time intervals. Furthermore, for an unsynchronized network, these range parameters are plagued with clock errors, which must be estimated and the respective clocks calibrated.


\subsection{Application} Our motivation for this work are \emph{inaccessible} mobile wireless networks, which have partial or no information of absolute co-ordinates and/or clock references. Such scenarios are prevalent in under-water communication \cite{chandrasekhar2006}, indoor positioning systems\cite{liu2007} and envisioned space based satellite networks with minimal ground segment capability. A particular project of interest is the Orbiting Low Frequency Antennas for Radio astronomy (OLFAR) \cite{rajanIEEEAero11}, a Dutch funded program which aims to design and develop a detailed system concept for an interferometric array of $\ge 10$ identical, scalable and autonomous satellites in space to be used as a scientific instrument for ultra low frequency observations ($0.3$MHz - $30$ MHz). The OLFAR cluster will be deployed far from the earth orbiting global positioning systems and hence cooperative network synchronization and localization is one of the key challenges in OLFAR, since no a priori information is available \cite{rajanIEEEAero13_1}. In comparison to the raw data exchange and the on board correlation in the satellites, the communication of measurements and proposed centralized algorithms have negligible impact, both in terms of communication and computational power.

\subsection{Contributions}  One of the main contributions in this paper is a novel joint time-range basis (Section \ref{sec:timeRange}), which combines the existing affine clock model (Section $\ref{sec:affineTime}$) with a generalized $(L-1)$th order non-linear range model (Section $\ref{sec:range}$) for an \emph{anchorless cluster of mobile nodes}. To the best of our knowledge, the two-way time stamp exchange between a pair of asynchronous nodes in motion has not been investigated before. In the presence of clock errors, the time varying distance measurements are corrupted with clock skews and clock offsets and the relation is addressed in Section $\ref{sec:range}$. The proposed joint basis is applied in a TWR framework and a \emph{Mobile Pairwise Least Squares (MPLS)} solution (Section \ref{sec:MGLS}) is proposed for a pair of mobile nodes, to estimate the clock skews, offsets and the range parameters of the pairwise distance between the nodes. Furthermore, for the entire network, all the clock skews, offsets and range parameters can be estimated using the proposed \emph{Mobile Global Pairwise Least Squares (MGLS)} algorithm (Section \ref{sec:MGLS}). More generally, when the order of distance approximation $L$ is unknown, iterative solutions are proposed for both the pairwise and global solutions.
A unique solution is non-existent without a constraint on the cost function (\eg clock reference node) and hence, a constrained framework is proposed. A new Constrained \Cramer\ Rao Bound (CCRB) (Section \ref{sec:CCRB}) is derived for the estimated clock and range parameters. In addition, instead of the classic constraint of using a single clock reference, an alternative \emph{sum constraint} is proposed (Section \ref{sec:constraintChoice}) based on an averaged clock reference, which is shown to yield about a factor better performance on the clock skew and offset estimation. The performance of the proposed algorithms and choice of constraints are analyzed using simulations (\ref{sec:simulations}).

\textit{Notation:} The element wise matrix Hadamard product is denoted by $\odot$, $(\cdot)^{\odot N}$ denotes element-wise matrix exponent and $\oslash$ indicates the element-wise Hadamard division. The Kronecker product is indicated by $\otimes$ and the transpose operator by ($\cdot)^T$. $\b1_N = [1, 1 \hdots ,1]^T, \bzero_N = [0, 0 \hdots, 0]^T \in \mathbb{R}^{N \times 1}$, are vectors of ones and zeros, respectively. $\bI_N$ is a $N \times N$ identity matrix, $\bzero_{M,N}$ is a $M \times N$ matrix of $0$, $\diag(\ba)$ represents a diagonal matrix with elements of vector $\ba$ on the diagonal and $\text{var}(\ba)$ denotes the corresponding variance.

\section{Joint time range basis} \label{sec:timeRange} \subsection{Affine time model} \label{sec:affineTime} Consider a network of $N$ nodes equipped with independent clock oscillators which, under ideal conditions, are synchronized to the global time. However, in reality, due to various oscillator imperfections and environment conditions the clocks vary independently and are inherently non-linear. Let $t_i$ be the local time at node $i$, then its divergence from the ideal \emph{true} time $t$ is to first order given by the affine clock model, \begin{eqnarray}
  t_{i} = \omega_{i}t+ \phi_i \quad \Leftrightarrow \quad\ \cC_{i}(t_i)\triangleq\ t=\ \alpha_{i}t_i +\beta_{i}
  \label{eq:affineTime}
\end{eqnarray} where $\omega_{i} \in \mathbb{R}_+$ and $\phi_i\in \mathbb{R}$ are the clock skew and clock offset of node
$i$ and the function $\cC_{i}(t_i)$ relates the local time $t_i$ to the \emph{true} time $t \triangleq\ \cC_i(t_i)$. In actuality, the clock skew ($\omega_i$) and clock offset ($\phi_i$) are time varying, but we assume they remain constant for small measurement time period (say $\Delta T$), which is often a reasonable assumption \cite{freris10}. Alternatively, the $2$nd part of (\ref{eq:affineTime}) shows the translation from local time $t_i$ to the global time $t$, where $[\alpha_i, \beta_i] \triangleq [\omega^{-1}_i,\ -\phi_i\omega^{-1}_i] $ are the calibration parameters needed to correct the local clock at node $i$. The clock skew and clock offset parameters for all $N$ nodes are represented by $\bomega=[\omega_1, \omega_2, \hdots, \omega_N]^T \in \mathbb{R}^{N \times 1}_+$ and $\bphi=[\phi_1, \phi_2, \hdots, \phi_N]^T \in \mathbb{R}^{N \times 1}$ respectively, and similarly the clock calibration parameters of the network are $\balpha \in \mathbb{R}^{N \times1}$ and $\bbeta \in \mathbb{R}^{N \times1}$. The unique relation between all the clock parameters is given by \begin{subequations}\label{eq:clockBasis}
  \begin{equation}
  \balpha   \triangleq \b1_{N}  \oslash \bomega \quad \Leftrightarrow \quad
  \bomega   \triangleq \b1_{N}  \oslash \balpha \label{subeq:alphaSkew}
  \end{equation}
  \begin{equation}
  \bbeta    \triangleq -\bphi   \oslash \bomega \quad \Leftrightarrow \quad
  \bphi     \triangleq -\bbeta  \oslash \balpha \label{subeq:betaOffset}
  \end{equation}
\end{subequations} Observe that for an ideal clock, $[\omega_i,\phi_i]=[1,0]$ immediately implies $[\alpha_i,\beta_i]=[1,0]$ and vice versa.


\subsection{Non-linear range model} \label{sec:range} In addition to clock variations, the nodes are also in motion with respect to each other.  Traditionally, when the nodes are fixed \cite{patwari05}, the pairwise propagation delay $\tau_{ij}$ between a node pair $(i,j)$ is $\tau_{ij} = c^{-1}d_{ij}$, where $d_{ij}$ is the fixed distance between the node pair and $c$ is the speed of the electromagnetic wave in the medium. \footnote{Without the loss of generality, we assume line of sight communication and hence all physical layer effects such as multi-path and shadowing are beyond the scope of this work. These scenarios can be addressed using existing techniques in literature \eg \cite{bellusci08}.} However, when the nodes are mobile, then the relative distances between the nodes are a non-linear function of time. For a small measurement time period $\Delta T$, the propagation delay $\tau_{ij}(t)$ between a node pair $(i,j)$ is then, classically a Taylor series, given by \begin{eqnarray}
\label{eq:rangeDefinition}
\tau_{ij}(t)  &\triangleq&  c^{-1} \cR_{ij}(t) \nonumber \\
              &\approx& c^{-1} (r^{(0)}_{ij} + r^{(1)}_{ij}t + r^{(2)}_{ij}t^2 +  \hdots  +r^{(L-1)}_{ij}t^{L-1} )
\end{eqnarray} where $\cR_{ij}(t)$ is the time varying pairwise distance between node pair $(i,j)$ and $\br_{ij}= \begin{bmatrix} r^{(0)}_{ij},r^{(1)}_{ij},r^{(2)}_{ij}, \hdots, r^{(L-1)}_{ij} \end{bmatrix}^T \in \mathbb{R}^{L-1 \times 1}$ contains all the range coefficients of the corresponding Taylor approximation. The order of approximation and the range of these $L$ coefficients depend on the initial position and the type of motion of the respective nodes. However, the propagation delay between the node pair is not measured at true time, instead by a local node clock, say node $i$. Hence, substituting the equation of ideal \emph{true} time $t$ from (\ref{eq:affineTime}), we have the propagation delay $\tau_{ij}(t_i)$ in terms of the local time $t_i$, \ie \begin{eqnarray}
\tau_{ij}(t_i)  &\triangleq&  c^{-1} \cR_{ij}(\cC_{i}(t_i)) \nonumber \\
                &\approx& \gamma^{(0)}_{ij} + \gamma^{(1)}_{ij}t_i + \gamma^{(2)}_{ij}t_i^2 + \hdots +\gamma^{(L-1)}_{ij}t_i^{L-1} \label{eq:rangeTranslation}
\end{eqnarray} where \begin{equation}
\cG_{ij}(t_i)= c^{-1} \cR_{ij}(\cC_{i}(t_i)) \label{eq:rangeTranslationFunction} \end{equation} describes the pairwise propagation delay \wrt the local time at $t_i$. The coefficients $\bgamma_{ij}= \begin{bmatrix} \gamma^{(0)}_{ij}, \gamma^{(1)}_{ij}, \gamma^{(2)}_{ij}, \hdots, \gamma^{(L-1)}_{ij}\end{bmatrix}^T \in \mathbb{R}^{L \times 1}$ are translated range parameters in terms of time, which incorporate the clock discrepancy of node $i$.

For the entire network, comprising of $\barN=\begin{pmatrix} N \\ 2 \end{pmatrix}$ \emph{unique} the pairwise links for $N$ nodes, all the unique range coefficients are given by \begin{eqnarray}
\label{eq:R}
\bR &=&
  \begin{bmatrix} \br_{12}, \br_{13}, \hdots, \br_{(N-1)N} \end{bmatrix}^T \in \mathbb{R}^{\barN \times L} \nonumber \\
  &=&
  \begin{bmatrix}
    r^{(0)}_{12}      & r^{(1)}_{12}      & \cdots & r^{(L-1)}_{12} \\
    r^{(0)}_{13}      & r^{(1)}_{13}      & \cdots & r^{(L-1)}_{13} \\
    \vdots            &                   & \ddots& \vdots      \\
    r^{(0)}_{(N-1)N}  & r^{(1)}_{(N-1)N}  & \cdots & r^{(L-1)}_{(N-1)N}\\
  \end{bmatrix}
\end{eqnarray} and along similar lines, we have the translated range coefficients \begin{eqnarray}
\label{eq:G}
\bGamma &=&
  \begin{bmatrix} \bgamma_{12}, \bgamma_{13}, \hdots, \bgamma_{(N-1)N} \end{bmatrix}^T \in \mathbb{R}^{\barN \times L}   \nonumber \\
&=&
  \begin{bmatrix}
    \gamma^{(0)}_{12}      & \gamma^{(1)}_{12}      & \cdots & \gamma^{(L-1)}_{12} \\
    \gamma^{(0)}_{13}      & \gamma^{(1)}_{13}      & \cdots & \gamma^{(L-1)}_{13} \\
    \vdots                 &                        & \ddots & \vdots      \\
    \gamma^{(0)}_{(N-1)N}  & \gamma^{(1)}_{(N-1)N}  & \cdots & \gamma^{(L-1)}_{(N-1)N}\\
  \end{bmatrix}
\end{eqnarray} where $r^{(l)}_{ij}$ and $\gamma^{(l)}_{ij}$ represent the unique $l$th order range coefficient for $(0 \le l \le L-1)$ of the node pair $(i,j)$ respectively. Furthermore, vectorizing these coefficient matrices, we have \begin{eqnarray} \label{eq:gammaRange_1}
\bgamma=\          \text{vec}(\bGamma) \in \mathbb{R}^{\bar{N}L \times 1}, \quad\
\br=\              \text{vec}(\bR) \in \mathbb{R}^{\bar{N}L \times 1}
\end{eqnarray}

Observe that although $\cG(\cdot)$ and $\cR(\cdot)$ are non-linear functions, $\cC_{i}(t_i)\ \forall\ 1\le i\le N$ is an affine translation and thus there exists a linear transformation matrix $\bG \in \mathbb{R}^{\bar{N}L \times \bar{N}L}$ containing $\begin{bmatrix} \balpha& \bbeta \end{bmatrix}$ such that \begin{eqnarray} \label{eq:rangeBasis} \br = \bG\bgamma &\Leftrightarrow&  \bgamma = \bG^{-1} \br. \end{eqnarray} The corresponding expression for $\bG$ is derived in Appendix \ref{ap:G}.

\subsection{Time range interrelation} \label{sec:Pairwise} In the following section we present a generalized TWR scenario where the joint time range basis is applied. Furthermore, an estimation process is described to obtain the the network parameters $\btheta=[\balpha, \bbeta, \bgamma]^T \in \mathbb{R}^{M \times 1}$ where $M=2N+\bar{N}L$, that are uniquely related to the desired unknown clock and range parameters $\bEta= [\bomega, \bphi, \br]^T \in \mathbb{R}^{M \times 1}$ by (\ref{eq:clockBasis}) and (\ref{eq:rangeBasis}) respectively. Finally the distance at discrete time intervals is obtained using (\ref{eq:rangeTranslation}).

%
%

\begin{figure}[tp] \centering
\includegraphics[scale=0.15]{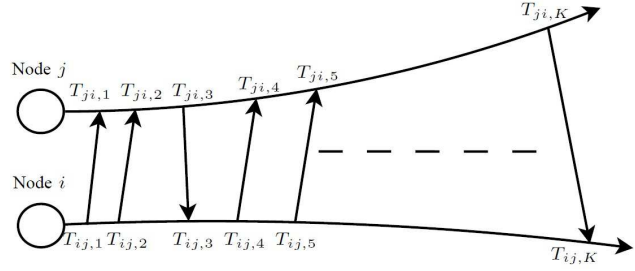}
\caption{ (\textit{Asynchronous pair of mobile nodes:}) A generalized Two Way Ranging (TWR) scenario between a pair of asynchronous nodes in motion, where the nodes exchange $K$ time stamps each. \emph{See Remark 1}.}
\label{fig:figPairWise}
\end{figure}

\section{Pairwise synchronization and ranging}
\subsection{Data Model} \label{sec:Datamodel} Consider a pair of mobile nodes $(i,j)$ with $i < j$, which are capable of two-way communication with each other as shown in \figurename\ \ref{fig:figPairWise}. The two nodes communicate messages back and forth, and the time of transmission and reception are  registered independently at respective nodes in respective local time coordinates. The $k$th time stamp recorded at node $i$ when communicating with node $j$ is denoted by $T_{ij,k}$ and similarly at node $j$ the time stamp is $T_{ji,k}$. Note that the total measurement period in this framework is $\Delta T= T_{ij,K}-T_{ij,1}$ seconds. The direction of the communication is indicated by $E_{ij,k}$, where $E_{ij,k} = +1$ for transmission from node $i$ to node $j$ and $E_{ij,k} = -1$ for transmission from node $j$ to node $i$. In contrast to previous cases of TWR \cite{ieee07,noh07,leng10} where the transmission and reception at a node was alternating, we do not presume any specific transmission/reception sequence\cite{rajanICASSP12,rajanCAMSAP11}. Furthermore, the propagation delay between the nodes at each time instant $1 \le k \le K$ is given by $\tau_{ij,k}= c^{-1}d_{ij,k}$, where $K$ is the number of time stamps recorded at each node \footnote{For the sake of simplicity, we assume the same K for all node pairs. The presented model can be easily generalized to different $K$ for each node pair within the network.} and $d_{ij,k}$ is the distance between the nodes at time instant $k$.

Under ideal circumstances, when the nodes are completely synchronized the noise free $k$th communication time markers are related as \begin{subnumcases}{T_{ji,k}=}
  T_{ij,k} + c^{-1}d_{ij,k} & for $i \rightarrow j$ \label{eq:idealUpLink}\\
  T_{ij,k} - c^{-1}d_{ij,k} & for $i \leftarrow  j$ \label{eq:idealDnLink}
\end{subnumcases} which can be combined as \begin{eqnarray}
T_{ji,k}  &=& T_{ij,k} + c^{-1}E_{ij,k}d_{ij,k}  \quad \quad \text{for}\ i \leftrightarrow j \\
          &=& T_{ij,k} + c^{-1}E_{ij,k}\cR_{ij}(T_{ij,k}) \label{eq:ieee2007_ideal}
\end{eqnarray} where the distance between the nodes $d_{ij,k}=c\tau_{ij,k}$ at time instant $k$ is $\cR_{ij}(T_{ij,k})$ defined in (\ref{eq:rangeDefinition}). However, due to clock uncertainties modeled in (\ref{eq:clockBasis}), and its subsequent influence on distance (\ref{eq:rangeTranslation}), (\ref{eq:ieee2007_ideal}) translates to \begin{eqnarray}
\label{eq:ieee2007_new}
\cC_i(T_{ij,k}) -\cC_j(T_{ji,k}) + E_{ij,k}\cG_{ij}(T_{ij,k})   &=& 0
\end{eqnarray} where without loss of generality, the time varying pairwise distance is expressed as a linear function of $t_i$ \ie time at node $i$.

\begin{figure*} 
\normalsize
\setcounter{eqnCounter1}{\value{equation}}
\begin{eqnarray}
\label{eq:basis}
\cC_i(T_{ij,k}+\eta_{i,k}) -\cC_j(T_{ji,k}+\eta_{j,k} ) + E_{ij,k}\cG_{ij}(T_{ij,k}+\eta_{i,k})   &=& 0  \\
\label{eq:basis1}
\alpha_iT_{ij,k} -\alpha_jT_{ji,k} +\beta_i -\beta_j + E_{ij,k}(\gamma^{(0)}_{ij} + \gamma^{(1)}_{ij}(T_{ij,k}+\eta_{i,k}) + \gamma^{(2)}_{ij}(T_{ij,k}+\eta_{i,k})^2 + \hdots\ )  &=& \alpha_j\eta_{j,k} - \alpha_i\eta_{i,k} \\
\label{eq:basis2}
\underbrace{\alpha_iT_{ij,k} -\alpha_jT_{ji,k} +\beta_i -\beta_j}_{\text{Clock parameters + Measurements}}   +\ \underbrace{E_{ij,k}}_{\text{Direction}} \underbrace{(\gamma^{(0)}_{ij} + \gamma^{(1)}_{ij}T_{ij,k} + \gamma^{(2)}T_{ij,k}^2 + \hdots\ )}_{\text{Range parameters + Measurements}} &=& \underbrace{\eta_{ij,k}}_{\text{noise}}
\end{eqnarray}
\setcounter{equation}{\value{eqnCounter1}}
\hrulefill
\vspace*{1pt}
\end{figure*}
\addtocounter{equation}{3}

Furthermore, in reality due to measurement noise on the time markers, (\ref{eq:ieee2007_new}) is (\ref{eq:basis}), where $\{\eta_{i,k}, \eta_{j,k}\}$ are noise variables plaguing the timing measurements at respective nodes. Rearranging the terms and incorporating the approximate range model for $\tau_{ij}(t_i)$ from ($\ref{eq:rangeTranslation}$) as a function of local time at node $i$ we have ($\ref{eq:basis1}$), which also includes the error due to Taylor series expansion. Expanding the equation and rearranging the terms begets ($\ref{eq:basis2}$), where $\eta_{ij,k}$ is the stochastic noise between the node pair $(i,j)$ at the $k$th instant.

\textit{ {\bf Remark \arabic{remarkCounter}}: (Mobility of the nodes during two-way communication):  In \figurename\ \ref{fig:figPairWise}, the curved lines symbolize the independent clock drifts in addition to the time varying distance between the nodes. In traditional TWR, for a fixed pair of nodes (\ie\ $L=1$), the pairwise distance $d_{ij,k}$ is assumed to be invariant for the total measurement period $\Delta T= T_{ij,K}-T_{ij,1}$. However, when the nodes are mobile, the distance at each time instance $k$ is dissimilar. Hence, instead of the classical assertion that the nodes are relatively stable over a time period $\Delta T$ \cite{ieee07,noh07,leng10}, we suppose that the nodes are relatively stable over a much smaller time period of $\delta t = |T_{ij,k} - T_{ji,k}|$ \ie the propagation time of the message. Furthermore, unlike previous cases \cite{ieee07,noh07,leng10} where the transmission and reception was alternating, the proposed setup imposes no pre-requisites on the sequence or number of two-way communications \cite{rajanCAMSAP11,rajanICASSP12,rajanEUSIPCO13}} \stepcounter{remarkCounter}

The curved lines symbolize the independent clock drifts in addition to the time varying distance between the nodes.  .


\subsection{Mobile Pairwise Least Squares (MPLS)} \label{sec:MPLS}
Extending ($\ref{eq:basis2}$) for all $K$ communications, a generalized joint clock and $(L-1)$th order range model for a pair of nodes is \begin{eqnarray}
\label{eq:basisMPLS}
\begin{bmatrix} \bA_{ij,1} & \bA_{ij,2} \end{bmatrix}
\begin{bmatrix} \alpha_{i} \\ \alpha_{j} \\ \beta_{i} \\ \beta_{j} \\ \gamma^{(0)}_{ij} \\ \gamma^{(1)}_{ij} \\ \gamma^{(2)}_{ij} \\ \vdots \\ \gamma^{(L-1)}_{ij} \end{bmatrix} = \bEta_{ij}
\end{eqnarray} where \begin{eqnarray}
\bA_{ij,1}&=&   \begin{bmatrix} \bt_{ij} & -\bt_{ji}  & \b1_{K}  &-\b1_{K} \end{bmatrix},  \\
\bA_{ij,2}&=&                   \bE_{ij}\bV_{ij}, \\
\bV_{ij}&=&     \begin{bmatrix} \bt^{\odot 0}_{ij}  & \bt^{\odot 1}_{ij} & \hdots & ,\bt^{\odot L-1}_{ij} \label{eq:pairwiseVandermonde} \end{bmatrix},
\end{eqnarray} contain the observation vectors  \begin{eqnarray}
\label{eq:timeVector}
\bt_{ij} &=&                            [T_{ij,1}, T_{ij,2}, \hdots, T_{ij,K}]^T \in \mathbb{R}^{K \times 1} \\
\label{eq:directionVector} \bE_{ij} &=& \diag(E_{ij,1},E_{ij,2}, \hdots, E_{ij,K}) \in \mathbb{R}^{K \times K}.
\end{eqnarray} The time markers recorded at node $i$ and node $j$ while communicating with each other are stored in $\bt_{ij}$ and $\bt_{ji}$ respectively, $\be_{ij}$ is a known vector indicating the transmission direction for each data packet and the noise vector $\bEta_{ij} \in \mathbb{R}^{K \times 1}$ is \begin{eqnarray} \label{eq:noise}
\bEta_{ij}  &=&     [\eta_{ij,1}, \eta_{ij,2}, \hdots, \eta_{ij,K}]^{T} \in \mathbb{R}^{K \times 1}.
\end{eqnarray}

Given a sufficiently large number of communications $K$ between the two nodes, the homogeneous system (\ref{eq:basisMPLS}) has a non-trivial solution spanning the null space of $[\bA_{ij,1}\ \bA_{ij,2}]$. The known Vandermonde matrix $\bV_{ij}$ is full rank for $K$ sufficiently large. Secondly, in $\bA_{ij,1}$ the column vectors $\b1_K$ and $-\b1_K$ are completely dependent and although $[\bt_{ij}\ -\bt_{ji}]$ is full rank, it is observed that the matrix $\bA_{ij,1}$ is rank deficient by $2$ and the corresponding null space is data dependent \cite{rajanCAMSAP11}.

A unique solution can be obtained by assuming either one of $\{\alpha_i, \alpha_j \}$ and either one of $\{\beta_i, \beta_j \}$ is known and thus eliminating respective columns in $\bA_{ij,1}$, which is in turn accomplished by choosing one of the two nodes as a clock reference\cite{rajanCAMSAP11}. More generally, we can translate the homogeneous equations into normal equations by asserting one of the two nodes as the reference node, say node $i$ with $[\alpha_i, \beta_i]=[1,0]$. This gives \begin{eqnarray} \label{eq:mpls_1} \bA_{ji}\btheta_{ij} &=&  \bb_{ij} + \bEta_{ij} \end{eqnarray} where \begin{eqnarray}
\label{eq:AijDefinition}
\bA_{ji}  &=&               [-\bt_{ji} \quad -\b1_{K} \quad  \bA_{ij,2}] \in \mathbb{R}^{K \times (L+2)}, \\
\label{eq:ThetaijDefinition}
\btheta_{ij} &=&            [\alpha_j \quad \beta_j \quad \bgamma^T_{ij}]^T \in \mathbb{R}^{(L+2) \times 1}, \\
\label{eq:bijDefinition}
\bb_{ij}&=&                   -\bt_{ij}.
\end{eqnarray} The Mobile Pairwise Least Squares (MPLS) solution is then obtained by minimizing the $l_2$ norm, \begin{eqnarray}
\label{eq:mplsSolution}
\Hat{\btheta}_{ij} =  \arg \min_{\btheta_{ij}} \; \|\bA_{ji}\btheta_{ij} - \bb_{ij} \|^2_2 =(\bA^T_{ji}\bA_{ji})^{-1}\bA^T_{ji}\bb_{ij} \end{eqnarray}  where $\Hat{\btheta}_{ij}= [\hat{\alpha}_j\; \hat{\beta}_j\; \hat{\bgamma}^T_{ij}]^T$ is an estimate of $\btheta$. Following, an estimate of the desired clock and range parameters $[\hat{\omega}_j\; \hat{\phi}_j\; \hat{\br}^T_{ij}]^T$ can then be obtained using $(\ref{eq:clockBasis})$ and $(\ref{eq:rangeBasis})$. An estimate of the approximated distance $d_{ij,k}$ between the nodes at the $k$th time instant is then from (\ref{eq:rangeTranslation}) \begin{equation}
\hat{d}_{ij,k} = c\Big(\hat{\gamma}^{(0)}_{ij} + \hat{\gamma}^{(1)}_{ij}T_{ij,k} + \hat{\gamma}^{(2)}_{ij}T^2_{ij,k} +  \hdots  +\hat{\gamma}^{(L)}_{ij}T^{L-1}_{ij,k} \Big)
\end{equation} and for all $1\le k \le K$, we have \begin{equation}
\label{eq:pairwiseDistanceApproxMPLS}
\widehat{\bd}_{ij} = c\bV_{ij}\widehat{\bgamma}_{ij}
\end{equation} where $\bV_{ij}$ is the Vandermonde matrix (\ref{eq:pairwiseVandermonde}) and $\hat{\bd}_{ij}=\begin{bmatrix} \hat{d}_{ij,1}, \hat{d}_{ij,2}, \hdots, \hat{d}_{ij,K} \end{bmatrix}^T  \in \mathbb{R}^{K \times 1}$ is the distance estimate between the node pair $(i,j)$ at all $K$ time instances.

More generally, when $L$ is unknown, solutions for increasing $L$ can be estimated using iterative MPLS (iMPLS) (based on order recursive least squares \cite{Kay1993}), which we briefly describe in Appendix \ref{sec:iMPLS} for the sake of completeness. This order recursive least squares not only implicitly estimates the unknown $L$ by incrementing the number of columns of the Vandermonde structure $\tilde{\bA}_{ij}$ iteratively, but also implements computationally economical updates of the inverse and solutions (\ref{eq:mplsSolution}).

\textit{ {\bf Remark \arabic{remarkCounter}}: (Feasibility of MPLS solution): The solution (\ref{eq:mplsSolution}) is feasible if $\bA_{ji,L} \in \mathbb{R}^{K \times (L +2)}$ is a square or tall matrix \ie the number of communications $K \ge (L +2)$. Secondly, to ensure full column rank, we require $\be_{ij} \ne -\b1_K$ and $\be_{ij} \ne +\b1_K$. In other words, among the $K \ge (L +2)$ data exchanges between the two nodes, there must be at least one transmission from $i$ to $j$ and $j$ to $i$ respectively.} \stepcounter{remarkCounter}

Although the MPLS solution is motivated for a mobile network of nodes, it is readily applicable for a network of immobile nodes. In that case, for a given node pair $\{i,j\}$ the estimated range parameter $r^{(0)}_{ij}$ indicates the fixed uncalibrated communication latency during the exchange of time stamps and the higher order range parameters indicate the latency fluctuations during communication.


\section{Network synchronization and ranging} \label{sec:network} 
\begin{figure}[tp] \centering
\includegraphics[scale=0.15]{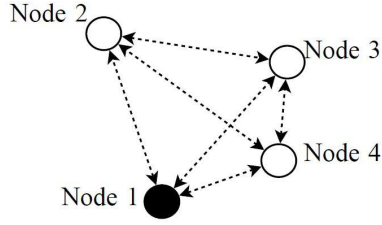} \caption{A network with $N=4$ nodes, each capable of two-way communication.
The clock skews and clock offsets of node 2, 3 and 4 are unknown and are to be estimated,  in addition to all unknown range parameters.}
\label{fig:figNetwork}
\end{figure}

We now extend the pairwise model in (\ref{eq:basisMPLS}) to the entire
network, \ie $N \ge 2$, and intend to find a global solution for joint ranging and synchronization. In the process, for the sake of notational simplicity we assume all nodes transmit $K$ messages, which is not mandatory. Secondly, we enforce the same approximation order on both time (first order) and distance (($L-1$)th order) for all node pairs (during the small measurement period). Thus, the proposed solution may not be accurate when the magnitude of the estimation parameters of some nodes vary eccentrically from the rest of the cluster within the approximation time period. As an illustration, \figurename\ \ref{fig:figNetwork} shows a network consisting of $N=4$ nodes with $\bar{N}=6$ pairwise communication links.

\subsection{Mobile Global Least Squares (MGLS)} \label{sec:MGLS} Aggregating (\ref{eq:basisMPLS}) for all pairwise links in the network, we have a linear global model of the form \begin{eqnarray}
\label{eq:dataModelMGLS}
\overbrace{[\bT \quad  \bH \quad \bar{\bV} ]}^{\large{\bA}}
\overbrace{\begin{bmatrix} \balpha \\ \bbeta \\ \bgamma \end{bmatrix}}^{\btheta} = \bEta
\end{eqnarray} where $\bar{\bV}=\bE\bV$ and $\bV \in \mathbb{R}^{\bar{N}K \times \bar{N}L}$ is a Vandermonde-like matrix given by\begin{equation} \label{eq:globalVandermonde}
\bV=              \begin{bmatrix} \bI_{\barN} \otimes \b1_K&  \bar{\bT}^{\odot 1}& \hdots& \bar{\bT}^{\odot L-1}\end{bmatrix}. \end{equation} $\bT\in \mathbb{R}^{\bar{N}K \times N}, \bar{\bT} \in \mathbb{R}^{\bar{N}K \times \barN}$ are measurement matrices contain the timing vectors recorded at all $N$ nodes. $\bH \in \mathbb{R}^{\bar{N}K \times N}$ is a matrix of $\pm\ \b1_K$ and $\bzero_K$, and $\bE \in \mathbb{R}^{\bar{N}K \times \barN L}$ contains all the direction vectors. The noise vector is represented as \begin{equation} \label{eq:globalNoise} \bEta = [\bEta_{12}^T, \bEta_{13}^T, \hdots, \bEta_{(N-1)(N)}^T]^T \in \mathbb{R}^{\bar{N}K \times 1} \end{equation} where each $\bEta_{ij}$ is given by (\ref{eq:noise}). We assume that the noise vectors for each pairwise communication $\bEta_{ij}$ are uncorrelated with one another, which may not be applicable for all communication schemes \eg broadcasting.


For $N=4$, $\bT$, $\bH$,  $\bar{\bT}$,  $\bE$ are of the form \begin{eqnarray}
\label{eq:defTE}
\bT &=&
\begin{bmatrix}
\bt_{12}& -\bt_{21} & &  \\
\bt_{13}& & -\bt_{31}&  \\
\bt_{14}& & & -\bt_{41} \\
 & \bt_{23}& -\bt_{32}& \\
 & \bt_{24}& & -\bt_{42} \\
 & &  \bt_{34} &-\bt_{43}\\
\end{bmatrix}, \nonumber \\
\bH &=&
\begin{bmatrix}
+\b1_{K} & -\b1_{K} & & \\
+\b1_{K} & & -\b1_{K}&  \\
+\b1_{K} & & & -\b1_{K} \\
 & +\b1_{K}& -\b1_{K}&  \\
 & +\b1_{K}& & -\b1_{K} \\
 & & +\b1_{K} & -\b1_{K}\\
\end{bmatrix}, \nonumber \\
\bar{\bT} &=& \diag(\bt_{12}, \bt_{13}, \bt_{14},\bt_{23}, \bt_{24}, \bt_{34}), \nonumber \\
\bE &=& \text{bdiag}(\bE_{12}, \bE_{13}, \bE_{14},\bE_{23}, \bE_{24}, \bE_{34}),
\end{eqnarray} where the empty spaces in matrices $\bT, \bH$ are entries with $0$. A similar structure can be obtained for $N \ge 4 $. The vector $\bt_{ij}$ contains the time stamps recorded at the $i$th node when communicating with the $j$th node in the network and is defined in $(\ref{eq:timeVector})$. Similarly, each vector $\be_{ij}$ contains the direction information of the corresponding pairwise communication and is defined in $(\ref{eq:directionVector})$.

Let us analyze the submatrices of $\bA$. We find $\bar{\bT}$ and $\bE$ are full column rank since they are block diagonal and subsequently, $\bar{\bV}= \bE\bV$ is a full rank matrix. $\bH$ is rank deficient by $1$, with a null space spanning $\{\b1_N\}$. The sparsely populated matrix $\bT$ containing the time stamp vectors is full rank. However, augmenting $\bT$ with the matrix $\bH$ further reduces the rank of $\bA$ by $1$ and hence we require at least $2$ constraints. This is expected, since a clock reference is needed to solve for unknown clock and range parameters of the network, as observed in Section \ref{sec:MPLS}.

\subsection{Equality Constrained Least Squares}\label{sec:ECLS} Traditionally, a simple constraint would be to choose a random node as the clock reference and thereby eliminating the rank deficiency in $\bA$. Following which, it is straightforward to formulate a global solution similar to (\ref{eq:mpls_1}), however in this section we will present a generic constrained least squares framework, the benefits of which will be discussed in Section \ref{sec:constrainedBenefits}.

Thus, more generally, the unknown vector $\btheta \in \mathbb{R}^{M \times 1}$, where $M= 2N+\bar{N}L$, can be estimated by minimizing the cost function \begin{eqnarray} \label{eq:mglsCostFunction}
\min_{\btheta}   && \|\ \bA\btheta\ \| ^2 \nonumber \\
\text{s.t.} && \bC\btheta = \bb
\end{eqnarray} where $\bA$ is the (rank-deficient) matrix defined in (\ref{eq:dataModelMGLS}), $\bC \in \mathbb{R}^{N_2 \times M}$ is a known constraint matrix and $\bb\in \mathbb{R}^{N_2 \times 1}$, where $N_2$ is the number of constraints. The equation $\bC\btheta=\bb$ implements the feasibility conditions, enforcing $N_2 \ge 2$ linearly independent constraints on $\btheta$. Assuming the constraints are selected such that $\begin{bmatrix} \bA \\ \bC \end{bmatrix} \in \mathbb{R}^{(\bar{N}K+N_2) \times L}$ is non singular and $\bb \ne \bzero_{N_2}$ \cite{lawson74}, the solution to $(\ref{eq:mglsCostFunction})$ is obtained by solving the Karush-Kuhn-Tucker (KKT) equations \cite{boydConvexOptimization} and is given by
\begin{equation}
\label{eq:mglsSolution}
\begin{bmatrix}
\hat{\btheta} \\
\hat{\blambda}
\end{bmatrix} =
\begin{bmatrix}
2\bA^T\bA & \bC^T \\
\bC      & \bzero_{N_2, N_2}\\
\end{bmatrix}^{-1}
\begin{bmatrix}
\bzero \\
\bb
\end{bmatrix}
\end{equation} where $\blambda \in \mathbb{R}^{N_2 \times 1}$ is the Lagrange vector. A detailed discussion on the choice of the constraint matrix $\bC$ is presented in Section \ref{sec:constraintChoice}.

Given the estimate $ \widehat{\btheta} =[\widehat{\balpha}^T, \widehat{\bbeta}^T, \widehat{\bgamma}^T ]^T$, an estimate of the clock parameters $\{\widehat{\bomega}, \widehat{\bphi}\}$ is estimated using (\ref{eq:clockBasis}) and the pairwise range parameters $\widehat{\br}$ between the nodes using (\ref{eq:rangeBasis}). Furthermore, all the \emph{unique} $\barN$ pairwise distances between the nodes $\widehat{\bd}= [\widehat{\bd}^T_{12},\widehat{\bd}^T_{13}, \hdots,\widehat{\bd}^T_{(N-1)N}]^T \in \mathbb{R}^{\bar{N}K \times 1}$ at all $K$ time instances are given by \begin{equation}
\label{eq:distanceEstimateMGLS}
\widehat{\bd}=  c\bV\widehat{\bgamma}
\end{equation} where $\bV$ is defined in $(\ref{eq:globalVandermonde})$. Similar to the iterative MPLS (iMPLS) solution (Appendix \ref{sec:iMPLS}), we propose an iterative equality constrained least squares algorithm (iMGLS) in Appendix \ref{sec:iMGLS} to estimate $\btheta$ in the presence of unknown $L$.

\begin{figure*}[tp]
\centering
\includegraphics[scale=0.22]{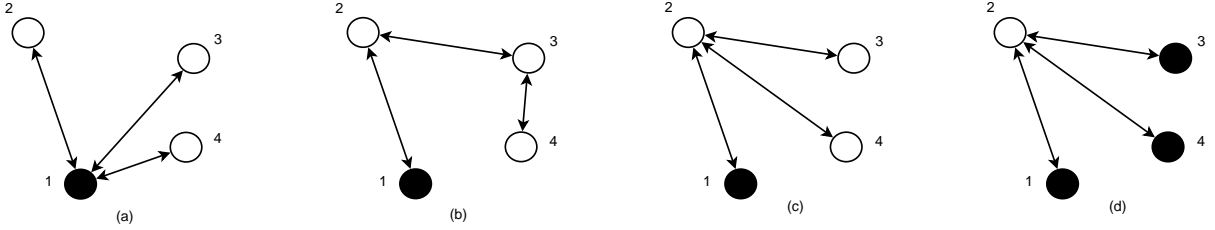} \caption{Four networks with $N=4$ nodes each capable of two-way communication. The node shaded in black is the clock reference. The 4 networks are illustrative examples where MGLS algorithm (and thus the constrained formulation) can be applied for network wide clock synchronization, despite missing communication links and multiple clock references.} \label{fig:figVariations}
\end{figure*}

\textit{ {\bf Remark \arabic{remarkCounter}}: (Extension to  partially connected networks): The closed form solution (\ref{eq:mglsSolution}) is for a full mesh network. More in general, if some pairwise communications links are missing then the corresponding rows in the primary matrix $\bA$ are dropped. Consequentially, the pairwise distances between those particular nodes cannot be estimated. However, despite missing links, network wide  synchronization is still feasible using the proposed algorithms if the primary matrix in (\ref{eq:mglsSolution}) is full rank \cite{rajanCAMSAP11,rajanICASSP12}. A few feasible topologies are illustrated in \figurename\ \ref{fig:figVariations}. For global synchronization, the network must consist of at least $N-1$ links, where every node has at least a single two-way communication link with one other node in the network.}\stepcounter{remarkCounter}

\textit{ {\bf Remark \arabic{remarkCounter}}: (Distributed MGLS): It is worth noting that, for $N=2$, the centralized MGLS is identical to the MPLS solution. However, the MGLS solution yields a more optimal estimate for the clock estimates (when $\bar{N} > N\ \ie N\ge4$) due to an increase in number of \emph{unique} pairwise links, which will be discussed in (\ref{sec:simulations}). Furthermore, although suboptimal, the MPLS is a distributed alternative to the centralized MGLS for estimating the clocks and range parameters. For large number of message exchanges and/or a large network of nodes \ie $K\barN \gg (L+2)$, the computational complexity of the MGLS algorithm is $\cO(KN^2L^2)$, which can be distributed efficiently using prevalent techniques \cite{bertrandDTLS11}. }\stepcounter{remarkCounter}

\section{Constrained \Cramer\ Rao Bounds} \label{sec:CCRB} \subsection{Noise modeling} In reality, the time markers in (\ref{eq:basis}) are plagued with measurement noise, which for simplicity is here assumed to be Gaussian \cite{Serpedin2009}. \footnote{Elsewhere, the noise on the time markers is also modeled as uniformly random variable (rising from quantization errors) or an exponential distribution \cite{ghaffar2002,Serpedin2009}.} Hence the noise on the nodes $\{i,j\}$ at the $k$th time instant in (\ref{eq:basis1}) are modeled as $\{\eta_{i,k}, \eta_{j,k} \} \sim \cN (0, 0.5\sigma^2)$, where without loss of generality, we assume the same noise variance on both transmission and reception markers. Subsequently, the cumulative noise vector $\eta_{ij,k}$ for the pairwise link (\ref{eq:basis2}), after ignoring the higher order noise terms, is \begin{equation} \eta_{ij,k}  = \alpha_j\eta_{j,k} - (\alpha_i + E_{ij,k}(\gamma^{(1)}_{ij}+ 2\gamma^{(2)}_{ij}T_{ij,k}+ \hdots\ ))\eta_{i,k} \end{equation} which is modeled as \begin{equation} \eta_{ij,k}  \sim\ \cN (0, 0.5\sigma^2(\alpha_j^2 + (\alpha_i+ \gamma^{(1)}_{ij}+ 2\gamma^{(2)}_{ij}T_{ij,k}+ \hdots\ )^2)). \end{equation} Note that the clock skews ${\omega_i}$ are typically very close to $1$ with errors of the order of $10^{-4}$ or so \cite{ieee07}. Hence, $\alpha_j^2 \approx\ 1\ \, \forall\ j \le N\ $ and such an approximation is satisfactory and is implicitly employed in various literature \cite{noh07,leng10,rajanCAMSAP11, wu11,wangTSP11,zheng10} for conventional fixed networks. Secondly, for $c= 3 \times 10^8 m/s$ the term $(\gamma^{(0)}_{ij}+ 2\gamma^{(1)}_{ij}T_{ij,k}+ \hdots\ )$ is scaled by $c^{-1}$ (by definition of $\bgamma$ in (\ref{eq:rangeijVectorDefinition}) and (\ref{eq:rangeBasisPairwise}) and thus is negligibly small for small measurement periods. Hence, the Gaussian noise is approximated to \begin{eqnarray} \eta_{ij,k}  &\sim& \cN (0, \sigma^2) \label{eq:noise_k} \end{eqnarray}

\textit{ {\bf Remark \arabic{remarkCounter}}: (Distance dependent noise): In reality, the pairwise noise $\eta_{ij,k}$ is also dependent on the distance between the nodes and the physical communication medium \cite{jia08}, in which case the noise is correlated with both channel effects and range parameters. The presented model can be readily extended to address these scenarios, where a weighted least square solution would be appropriate in contrast to the proposed least squares solution.}\stepcounter{remarkCounter}

\subsection{Lower Bounds for joint time-range estimation} In order to verify the performance of the proposed algorithms, we derive a Constrained \Cramer\ Rao lower Bound (CCRB) for the joint affine clock and $L-1$th order range model defined in (\ref{eq:dataModelMGLS}). The error vector $\bEta$ in (\ref{eq:dataModelMGLS}) is Gaussian by assumption and following immediately, the Constrained \Cramer\ Rao Bound (CCRB) on the error variance for an unbiased estimator is given by \cite{stoica1998} \begin{eqnarray}
\label{eq:crbMGLS} {\mathbb{E}} \left \{ (\hat{\btheta}-\btheta)(\hat{\btheta}-\btheta)^T \right \} \ge \cSigma_{\theta}
&\triangleq& \begin{bmatrix} \bSigma_{\alpha}& * & *\\ * & \bSigma_{\beta} & * \\ * & *& \bSigma_{\gamma} \end{bmatrix} \nonumber \\
&=& \bU(\bU^T\bF\bU)^{-1}\bU^T \end{eqnarray} where $\cSigma_{\theta}$ is the \Cramer\ Rao lower Bound on  $\btheta = \begin{bmatrix} \balpha & \bbeta & \bgamma \end{bmatrix}$, $*$ represent entries not of interest, $\bU \in \mathbb{R}^{M \times (M-N_2)}$ with $M=2N+\bar{N}L$ is an orthonormal basis for the null space of the constraint matrix $\bC$ with $N_2$ constraints, and
\begin{equation}
\bF = \sigma^{-2}\bA^T\bA= \left[
        \begin{array}{c c c}
        \bT^T\bT         & \bT^T\bH       & \bT^T\bar{\bV} \\
        \bH^T\bT         & \bH^T\bH       & \bH^T\bar{\bV}  \\ 
        \bar{\bV}^T\bT  & \bar{\bV}^T\bH & \bar{\bV}^T\bar{\bV} \\
        \end{array} \right] \in \mathbb{R}^{M \times M},
\label{eq:FIM}
\end{equation}
is the Fisher Information Matrix (FIM). Moreover, since the system parameters $\bEta=  \begin{bmatrix}\bomega &  \bphi& \br \end{bmatrix}$ can be uniquely derived from $\btheta$, we have the CRB on the estimates of $\bEta$ from standard error propagation formulas \cite{Kay1993} as, \begin{equation} \label{eq:crbSigmaEta} \cSigma_{\eta} \triangleq\  \begin{bmatrix} \bSigma_{\omega}& * & *\\ * & \bSigma_{\phi} & * \\ * & *& \bSigma_{r} \end{bmatrix} =\ \bJ_{\theta{\eta}}\bSigma_{\theta}\bJ^T_{\theta{\eta}} \end{equation} where $\bSigma_{\theta}$ is given by ($\ref{eq:crbMGLS}$) and $\bJ_{\theta{\eta}} \in \mathbb{R}^{M \times M} $ is the Jacobian of the transformation of $\bEta$ from $\btheta$ (Appendix \ref{sec:appendixJacobian}). Following immediately, given the lower bound on the variance of $\bgamma$ as $\bSigma_{\gamma} \in \mathbb{R}^{\bar{N}L \times \bar{N}L}$, the lower bound on the variance of the distance estimate (\ref{eq:distanceEstimateMGLS}) is \begin{equation} \label{eq:crbDistance} \cSigma_{d} =\  c^2\bV\bSigma_{\gamma}\bV^T \end{equation} where $\bV$ is the Vandermonde-like matrix (\ref{eq:globalVandermonde}).

\textit{ {\bf Remark \arabic{remarkCounter}}: (Generalization of MGLS, CCRB): The global solutions namely, Global Least Squares (GLS)\cite{rajanCAMSAP11} , Extended Global Least Squares (EGLS) \cite{rajanICASSP12}, \text{Extended$^2$} Global Least Squares \EEGLS \cite{rajanEUSIPCO13} (and corresponding pairwise solutions \{PLS, EPLS, \EEPLS\}) are special cases of MGLS (and MPLS) for the distance approximation of $L= 1,2,3$ respectively. In addition, the choice of range approximation order $L$ is automatically estimated using the proposed iterative solutions (iMGLS, iMPLS). Similarly, the new CCRB (\ref{eq:crbMGLS}) and the Jacobian $\bJ_{\theta\eta}$ (\ref{eq:JacobianMGLS}) are also generalizations of the respective lower order models proposed in \cite{rajanCAMSAP11,rajanICASSP12,rajanEUSIPCO13} for any $L \ge 1$.}\stepcounter{remarkCounter}

\section{On the choice of clock reference} \label{sec:constraintChoice}  Observe that the solution to $\btheta$ in (\ref{eq:mglsSolution}) and its corresponding performance (\ref{eq:crbMGLS}), (\ref{eq:crbSigmaEta}) is not only data dependent, but also depends on the choice of constraints. The primary matrix $\bA$ is rank deficient by $2$ and hence, $N_2 \ge 2$  feasible constraints are needed on the clock parameters to ensure a unique solution in (\ref{eq:mglsSolution}). In view of achieving an optimal solution, we discuss three potential constraints, namely (a) the classic constraint, (b) a nullspace constraint and (c) the sum constraint.

\subsection{Classic constraint} The minimum requirement for a feasible solution is to use an arbitrary node $i$ as a clock reference, \ie the constraint $\alpha_i=1$ and $\beta_i=0$, which yields the \emph{classic constraint}, \begin{eqnarray} \label{eq:classicConstraint}
\bC_1 = \left[ \begin{array}{c|c|c}
\bc_i^T       & \bzero^T_{N}  & \bzero^T_{\bar{N}L}\vspace{1mm} \\
\bzero^T_{N}  & \bc_i^T       & \bzero^T_{\bar{N}L}
\end{array} \right] , \quad
\bb_1 = \begin{bmatrix} 1 \\ 0 \end{bmatrix}.
\end{eqnarray} where \begin{equation}
\label{eq:c_i}
\bc_i = [\bzero^T_{i-1} ,\ 1,\ \bzero^T_{N-i}]^T \quad\ \in \mathbb{R}^{N \times 1}.
\end{equation} Such a constraint is often utilized without further discussion for clock synchronization in a network of fixed nodes \cite{ieee07,Serpedin2009,wu11} and much of the literature on localization \cite{zheng10}.

\subsection{Nullspace constraint} Among the set of all feasible linearly independent constraints, the pseudo-inverse of the unconstrained FIM yields the lowest value for the total variance on all estimated parameters \cite{carvalho2000}. Let the spectral decomposition of the rank deficient FIM be \begin{equation}
\bF= [\bV_1 \quad \bV_2] \begin{bmatrix} \bLambda_1& \bzero \\ \bzero & \bLambda_2 \end{bmatrix}[\bV_1 \quad \bV_2]^T
\approx\ \bV_1\bLambda_1\bV^T_1\label{eq:FIMevd} \end{equation} where $\bLambda_1$ is a diagonal matrix containing the non-zero eigenvalues and $\bV_1$ the corresponding eigenvectors. Now, let $\bC_2$ be the nullspace constraint matrix such that the range of $\bC^T_2$ spans the null space of $\bF$ (\ie in the range of $\bV_2$). Subsequently, the orthogonal basis for the null space of $\bC_2$ \ie $\bU_2$ spans the range of $\bV_1$, and the trace of the CCRB (\ref{eq:crbMGLS}) is \begin{eqnarray}
\text{Tr}\left( \bSigma_{\theta} \right)
  &=& \text{Tr}\left[ \bU_2(\bU_2^T\bF\bU_2)^{-1}\bU_2^T\right] \nonumber \\
  &=& \text{Tr}\left[ \bV_1(\bV_1^T(\bV_1\bLambda_1\bV^T_1)\bV_1)^{-1}\bV_1^T\right] \nonumber \\
  &=& \text{Tr}\left[  \bLambda^{-1}_1\right] =\ \text{Tr}\left[  \bF^{\dagger} \right]
  \label{eq:FIMtrace}
\end{eqnarray} where we use the property $\bV^T_1\bV_1= \bI$ and exploit the cyclic nature of the trace operator. Hence, the nullspace constraint yields the pseudo-inverse of the unconstrained FIM, which is the lowest achievable total variance on all estimated parameters. This implies that any set of vectors which span the nullspace of the FIM form an optimal constraint for the system. However, note that while the nullspace constraint guarantees the lowest variance on $\btheta$, it offers little insight on the optimality of the independent parameters $\balpha, \bbeta, \bgamma$ and subsequently on the translated parameters of interest $\bomega, \bphi$ and $\bd$. Furthermore, this constraint is data dependent and presents no physical intuition on the estimated parameters.

\subsection{Sum constraint} In the pursuit of a data independent constraint and inspired by \cite{wijnholdsConstrained06}, we propose a \emph{sum constraint}, whereby we enforce \emph{the sum of all $\alpha_i$ to be $1$ and the sum of all $\beta_i$ to be 0, \ie $\bSum^N_i \alpha_i=1$ and $\bSum^N_i \beta_i=0$}, which begets a new constraint matrix \begin{eqnarray}
\label{eq:sumConstraint}
\bC_2 =
\left[
\begin{array}{c|c|c}
\b1^T_{N}     & \bzero^T_{N} & \bzero^T_{\bar{N}L}\\
\bzero^T_{N}  & \b1^T_{N}    & \bzero^T_{\bar{N}L}
\end{array} \right] , \quad
\bb_2 = \begin{bmatrix} 1 \\ 0 \end{bmatrix}.
\end{eqnarray} The \emph{sum constraint} proposes a virtual ``average'' clock, which in turn is governed by the clock errors $\{\balpha, \bbeta\}$ of all the clocks in the network and thereby alleviates a single clock reference which maybe potentially unstable. In case of the classic constraint with a single clock reference, the variance of the reference clock parameters is artificially put to zero and thereby accruing its variance to all other clock parameter estimates. In comparison, the sum constraint computes the average $\beta_i$ (and $\alpha_i$) for all the nodes, which leads to about a factor $2$ reduction in the variance of the estimate of $\beta_i$ (and $\alpha_i$) \cite{wijnholdsConstrained06}, and subsequent improvement on $\bomega$ and $\bphi$ due to averaging, as observed in the simulations (Section \ref{sec:simulations}).

As shown in (\ref{eq:FIMtrace}), any set of constraints that span the null space of the FIM yield an optimal estimate of the unknown parameter. Among the pair of proposed sum constraints on $\balpha, \bbeta$, observe that the second constraint $[\bzero^T_{N}\; \b1^T_{N}\;  \bzero^T_{\bar{N}L}]^T$ indeed lies in the null space of the FIM (\ref{eq:FIM}), since $\bH \b1_N = \bzero_{\bar{N}}$. However, a similar argument cannot be made for the constraint on $\balpha$, \ie $[\b1^T_{N}\; \bzero^T_{N}\;  \bzero^T_{\bar{N}L}]^T$, thus the sum  constraint is not yet optimal (unlike the case in \cite{wijnholdsConstrained06}), although it is seen to be close to optimum in simulations.

\subsection{Benefits of the constrained formulation} \label{sec:constrainedBenefits} Contrary to the pairwise algorithm MPLS, which was formulated as a least square solution, the global algorithm is structured as a constrained least squares problem. Such a \emph{generic framework} enables the user to incorporate additional \textit{a priori} information into the constraint matrix $\bC$ and thereby obtain a lower variance on the clock and range estimates. For example, if the network has three reference nodes, say node 1, 3, and 4, which is common in joint TOA localization and synchronization \cite{zheng10,zhu10} (refer \figurename\ \ref{fig:figVariations}(d)), then by increasing the number of rows $M_2$ of the constraint matrix $\bC$, such as  \begin{eqnarray}
\label{eq:constraintExample_1}
\acute{\bC}=
\left[
\begin{array}{c|c|c}
\bc^T_1       & \bzero^T_{N}  & \bzero^T_{\bar{N}L}\\
\bzero^T_{N}  & \bc^T_1       & \bzero^T_{\bar{N}L}\\
\bc^T_3       & \bzero^T_{N}  & \bzero^T_{\bar{N}L}\\
\bzero^T_{N}  & \bc^T_3       & \bzero^T_{\bar{N}L}\\
\bc^T_4       & \bzero^T_{N}  & \bzero^T_{\bar{N}L}\\
\bzero^T_{N}  & \bc^T_4       & \bzero^T_{\bar{N}L}
\end{array} \right] , \quad
\acute{\bb} = \begin{bmatrix} \alpha_1 \\ \beta_1 \\ \alpha_3 \\ \beta_3 \\ \alpha_4 \\ \beta_4 \end{bmatrix}
\end{eqnarray} a more optimal estimate can be obtained for the unknown clock parameters of node $2$. As a special case, if there are one-way communication links from the reference nodes to node $2$ and the reference nodes directly communicate their \emph{true} time, then \figurename\ \ref{fig:figVariations}(d) simplifies to the conventional GPS based synchronization and ranging\cite{kaplan06}. Likewise, for $L=1$, in a network with adequate known node positions, one can incorporate known pairwise distances in the constraint matrix to yield higher accuracy in overall estimates. The formulation in (\ref{eq:mglsCostFunction}) is thus a convenient framework to incorporate various prevalent scenarios.

\begin{figure*} \centering
  \begin{subfigure}[b]{0.32\textwidth}
    \centering
    \includegraphics[scale=0.35]{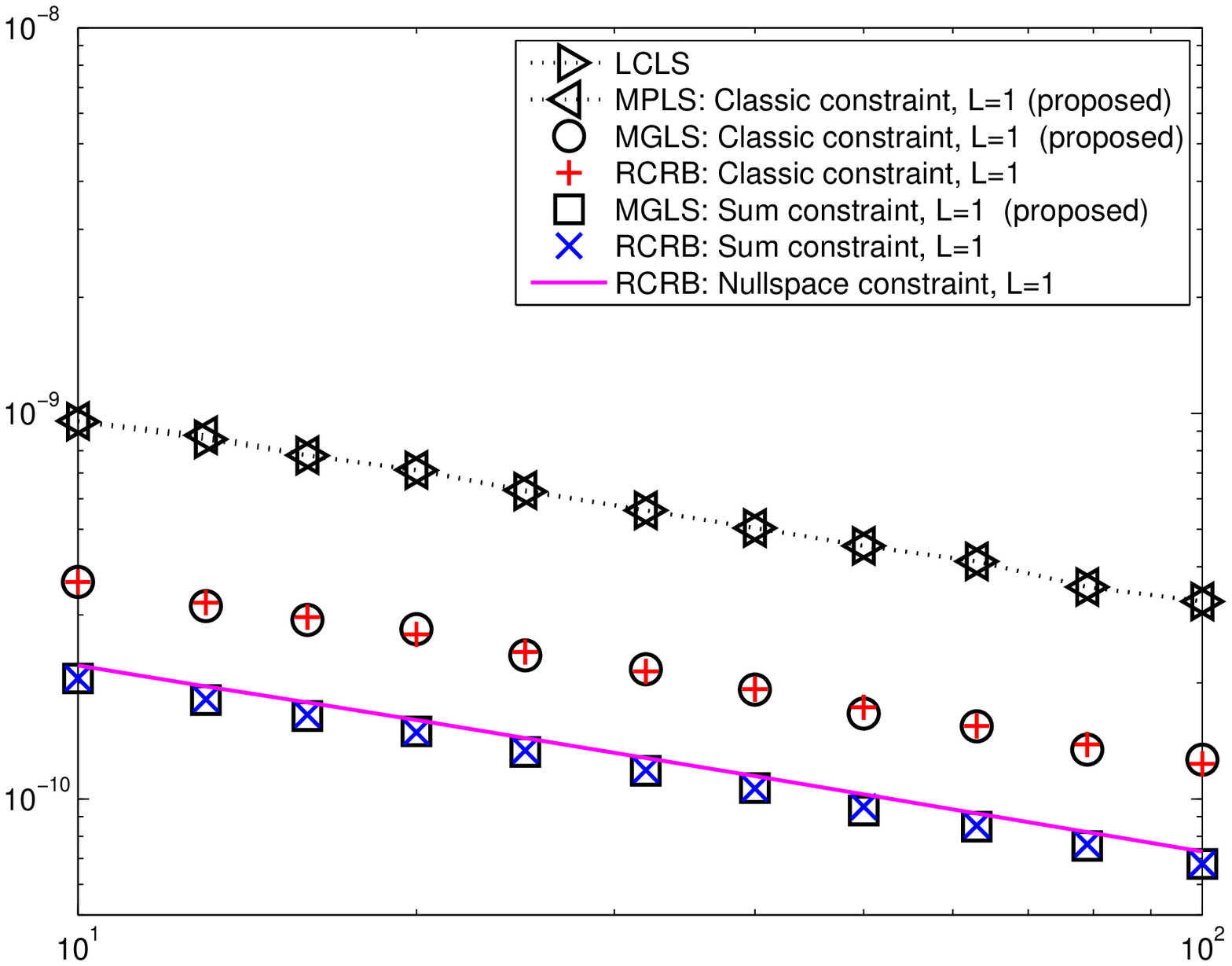}
    \rput(0.5,+.25){\tiny{Number of two-way communications (K)}}
    \rput(0.4, 5.2){\small{RMSE of clock skew ($\hat{\bomega}$)}}
    \caption{}
  \end{subfigure}
  \begin{subfigure}[b]{0.32\textwidth}
    \centering
    \includegraphics[scale=0.35]{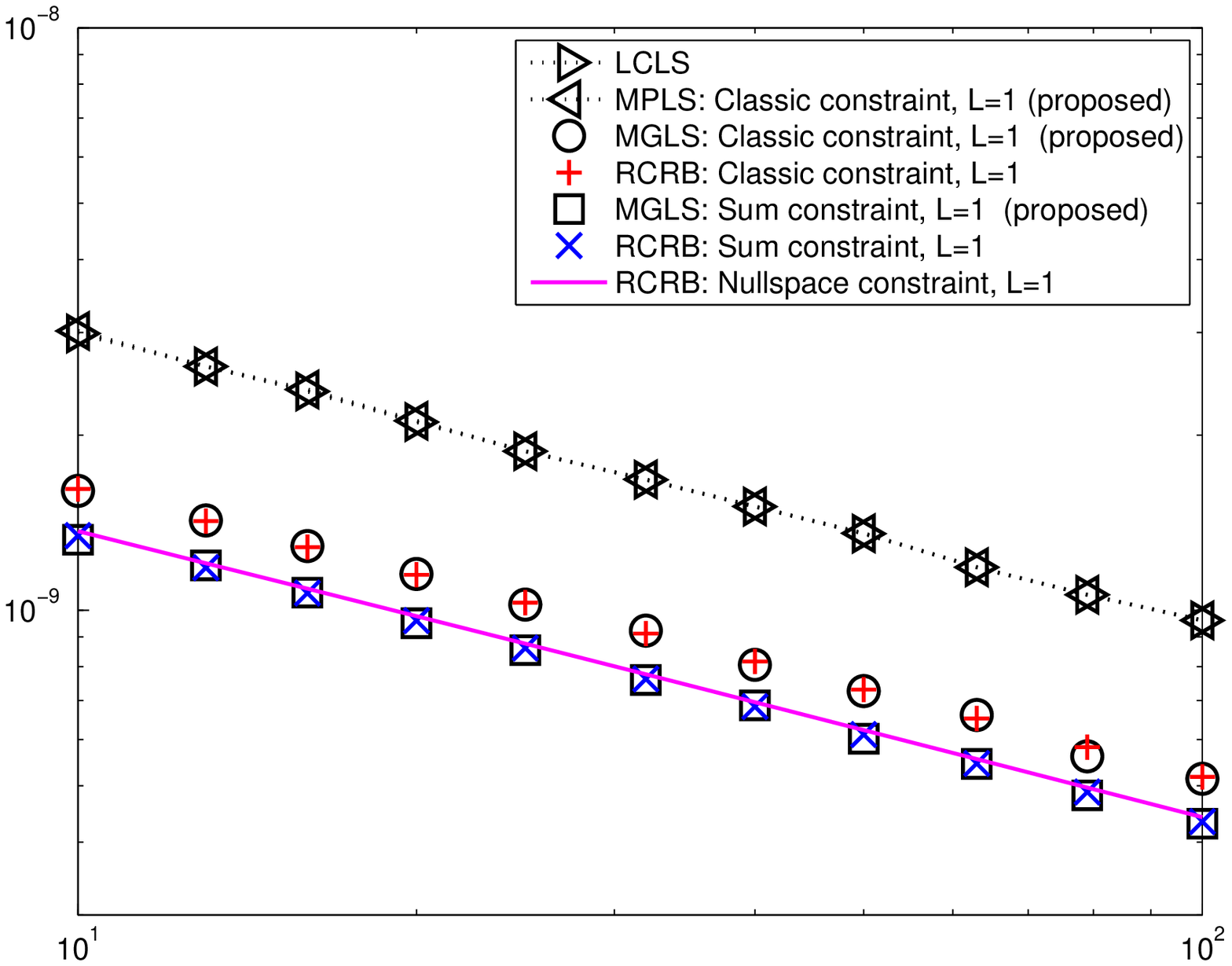}
    \rput(0.5,+.25){\tiny{Number of two-way communications (K)}}
    \rput(0.4, 5.2){\small{RMSE of clock offset ($\hat{\bphi}$)}}
    \caption{}
  \end{subfigure}
  \begin{subfigure}[b]{0.32\textwidth}
    \centering
    \includegraphics[scale=0.35]{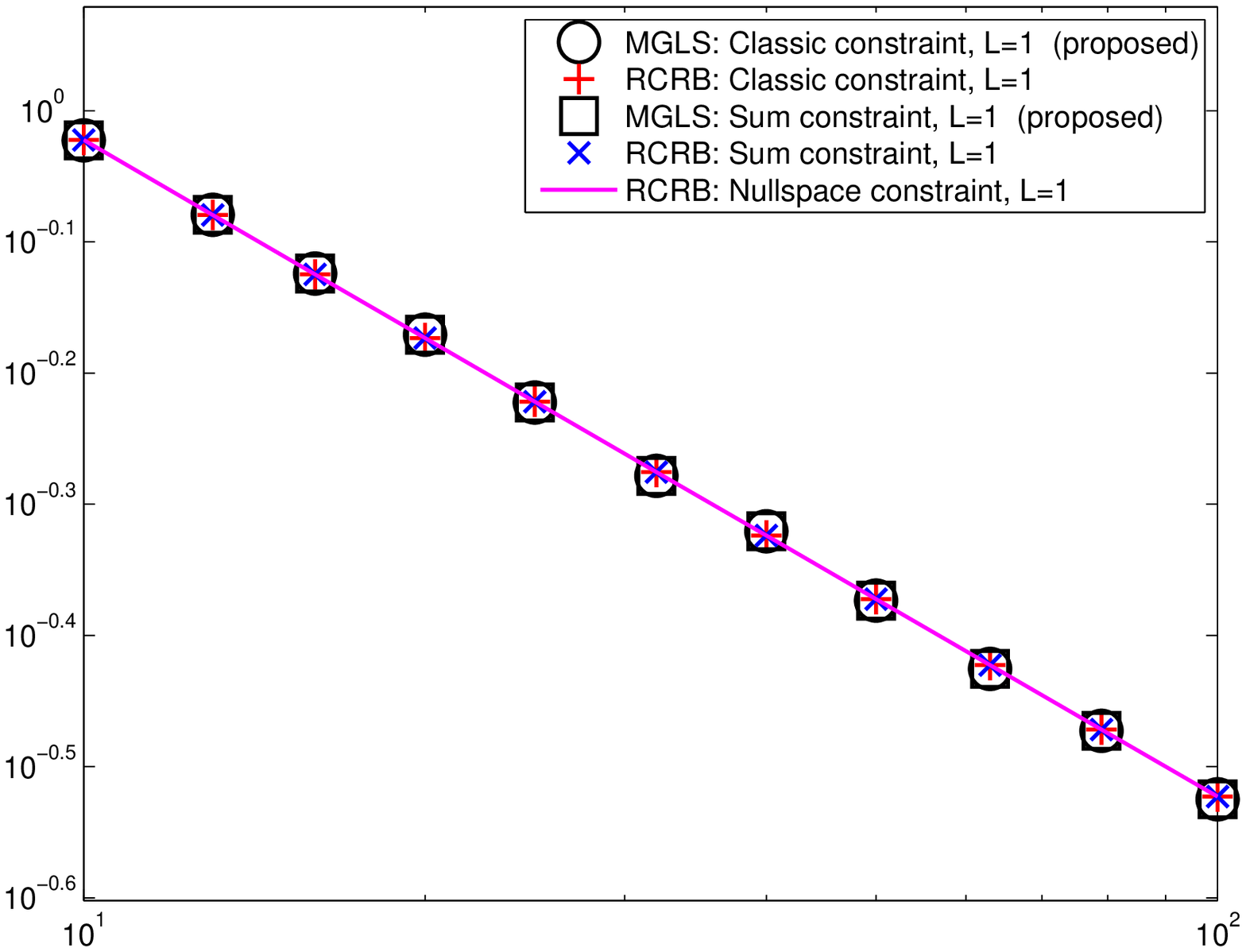}
    \rput(0.5,+.25){\tiny{Number of two-way communications (K)}}
    \rput(0.4, 5.2){\small{RMSE of distance ($\hat{\bd}$)}}
    \caption{}
  \end{subfigure}
  \caption{\small \emph{\textbf{Immobile network: Varying $K$:}} RMSEs (and RCRBs) of (a) clock skew, (b) clock offset and (c) distances for varying number of communications ($K$) between the $N=10$ \emph{fixed} nodes for $\sigma=10^{-8}$ seconds}
  \label{fig:staticK}
\end{figure*}


\begin{figure*} \centering
  \begin{subfigure}[b]{0.32\textwidth}
    \centering
    \includegraphics[scale=0.35]{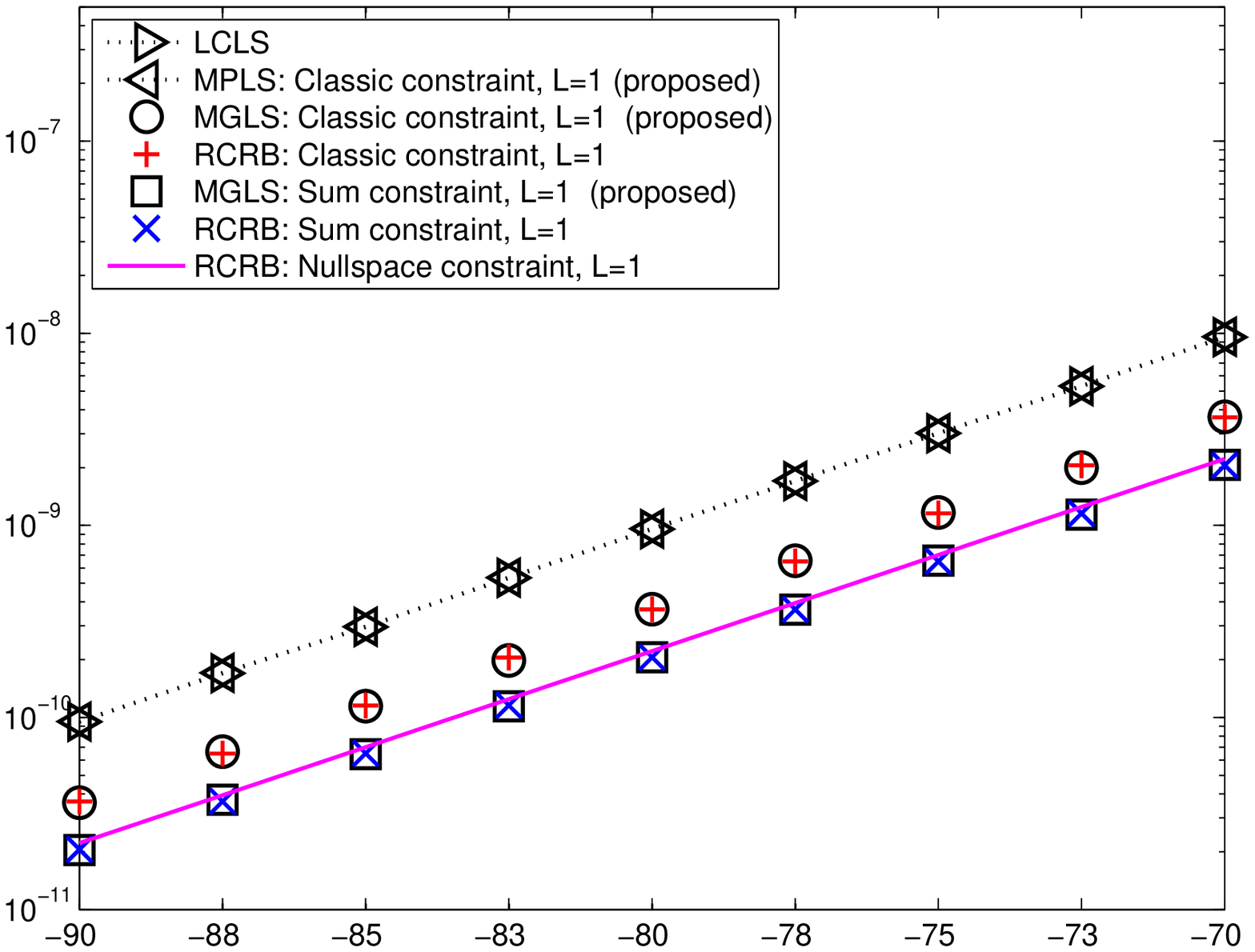}
    \rput(0.5,+.25){\tiny{$10\log_{10}(\sigma)$ [dB second]}}
    \rput(0.4, 5.2){\small{RMSE of clock skew ($\hat{\bomega}$)}}
    \caption{}
  \end{subfigure}
  \begin{subfigure}[b]{0.32\textwidth}
    \centering
    \includegraphics[scale=0.35]{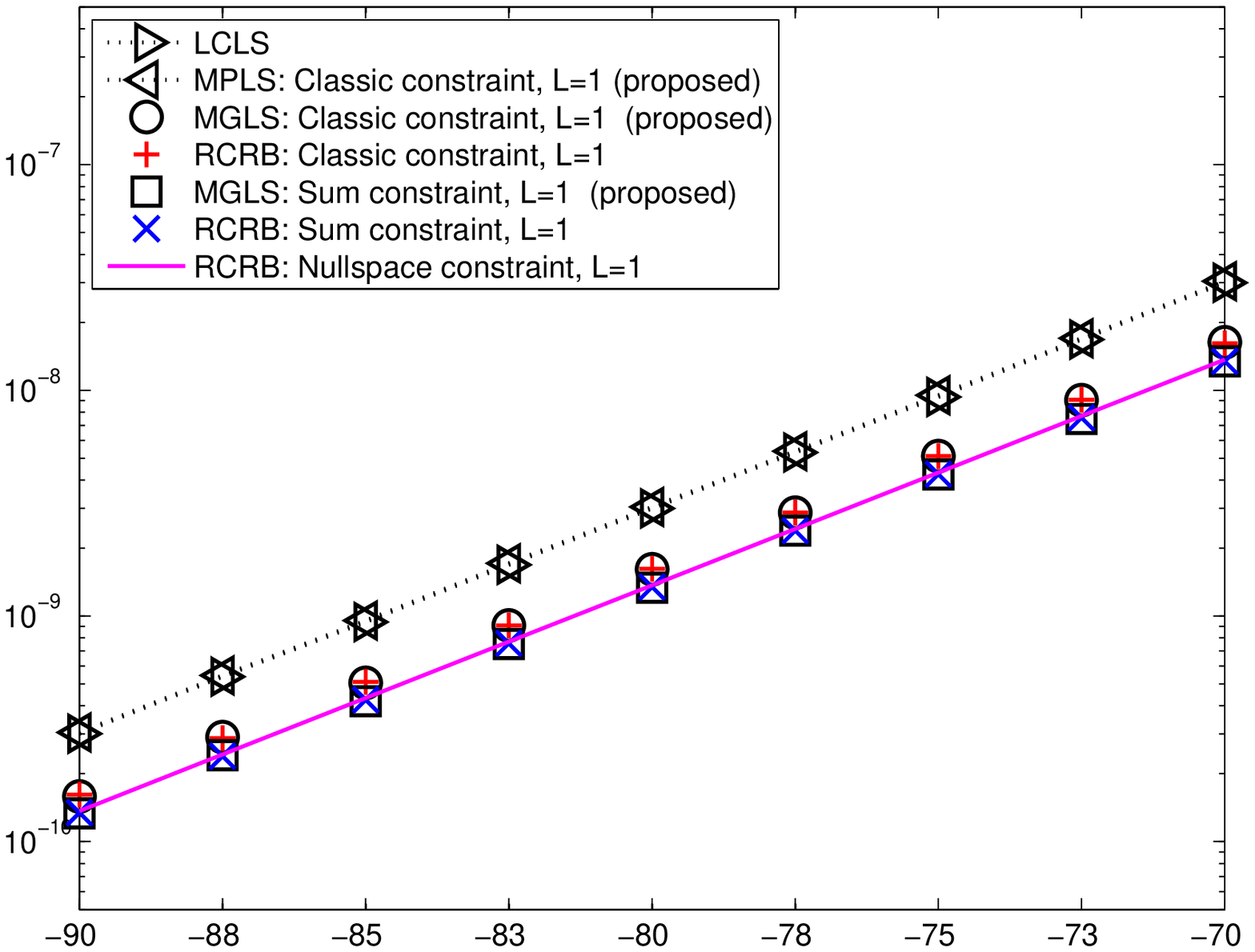}
    \rput(0.5,+.25){\tiny{$10\log_{10}(\sigma)$ [dB second]}}
    \rput(0.4, 5.2){\small{RMSE of clock offset ($\hat{\bphi}$)}}
    \caption{}
  \end{subfigure}
  \begin{subfigure}[b]{0.32\textwidth}
    \centering
    \includegraphics[scale=0.35]{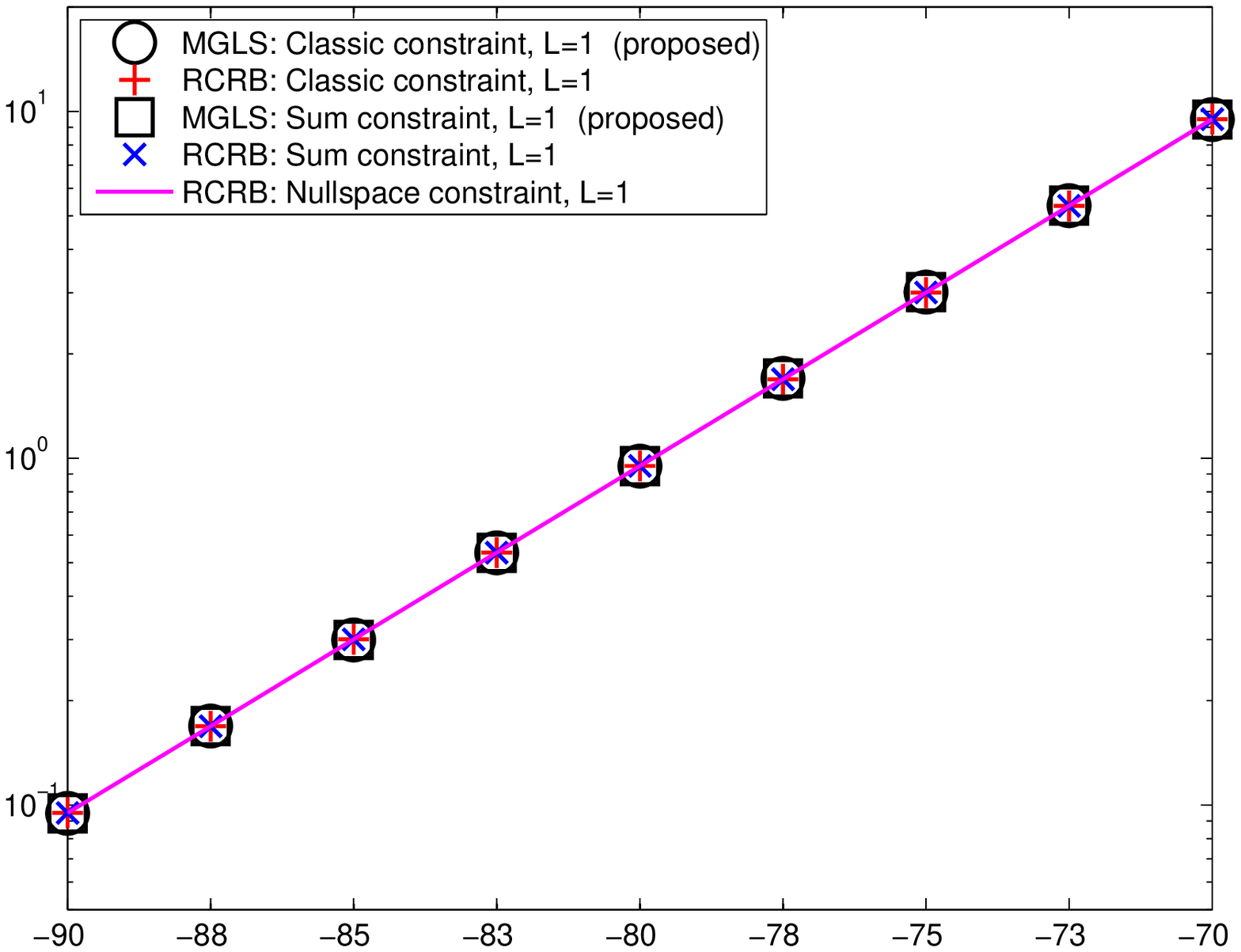}
    \rput(0.5,+.25){\tiny{$10\log_{10}(\sigma)$ [dB second]}}
    \rput(0.4, 5.2){\small{RMSE of distance ($\hat{\bd}$)}}
    \caption{}
  \end{subfigure}
  \caption{\small \emph{\textbf{Immobile network: Varying $\sigma$:}} RMSEs (and RCRBs) of (a) clock skew, (b) clock offset and (c) distances for a cluster of \emph{fixed} nodes, for varying noise ($\sigma$) on the time measurements with $K=20$ number of communication}
  \label{fig:staticSNR}
\end{figure*}


\section{Simulations} \label{sec:simulations} Simulations are conducted to evaluate the performance of the proposed estimators. We consider a network of $N=10$ mobile nodes, each capable of two-way communication with each other. The nodes transmit and receive time stamps alternatingly and thus the direction matrix $\bE$ is (\ref{eq:defTE}), where $\be_{ij}= [+1, -1]^{T} \otimes \b1_{0.5K}$. The transmission time markers $\bt_{ij}$ are linearly distributed within a small measurement time interval of $\Delta T= [-5, 5]$ seconds. All the nodes are equipped with independent clock oscillators, whose clock skews ($\bomega$) and clock offsets ($\bphi$) are uniform randomly distributed in the range $[1-10 \text{ppm},1+10\text{ppm}]$ and $[-10,+10]$ seconds respectively, which are given by (\ref{eq:skewValues}) and (\ref{eq:offsetValues}) respectively.

\begin{figure*} [!b] \normalsize
\hrulefill
\normalsize
\setcounter{eqnCounter3}{\value{equation}}
\begin{eqnarray}
\label{eq:positionValues}
\bX&=&
  \begin{bmatrix}
   615&  -764&   -19&   823&   899&  -894&   994&   597&  -931&   815 \\
  -130&   443&   296&  -973&  -178&  -770&  -757&   782&   780&  -445
  \end{bmatrix} \\
\label{eq:velocityValues}
\dot{\bX}&=&
  \begin{bmatrix}
    -7&    -5&    -3&    -7&    -7&     4&    -4&     6&     7&     1 \\
     9&     4&     4&    -9&   -10&     1&    -8&    -4&     7&     9
  \end{bmatrix} \\
\label{eq:skewValues}
\bomega&=&
  \begin{bmatrix}1.0000& 0.9999& 0.9994& 1.0005& 1.0001& 0.9999& 1.0009& 1.0000& 0.9997& 1.0006   \end{bmatrix}^T \\
\label{eq:offsetValues}
\bphi &=&
  \begin{bmatrix} 0& 9.4215& 6.9275& 0.1200& -4.4225& 4.9323& -5.2614& 9.1469& 2.4052& 2.0052 \end{bmatrix}^T
\end{eqnarray}
\setcounter{equation}{\value{eqnCounter3}}
\vspace*{1pt}
\end{figure*}
\addtocounter{equation}{2}

The metric used to evaluate the performance of the estimators is the Root Mean Square Error (RMSE) given by $\text{RMSE} (\hat{\bz},\bz) = \sqrt{N_{\text{exp}}^{-1}\sum^{N_{\text{exp}}}_{n=1}|| \hat{\bz}(n)-\bz ||^2}$, where $\hat{\bz}(n)$ is the $n$th estimate of the unknown vector $\bz \in \mathbb{R}^{N \times 1}$ to be estimated and the number of experiments is $N_{\text{exp}}=1000$.  Furthermore, along with the RMSE plots, the square Root of the constrained \Cramer\ Rao Bounds (RCRB) derived in Section \ref{sec:CCRB} are also plotted for the three constraints discussed in Section \ref{sec:constraintChoice}. In case of the classic constraint, node $1$ is assumed to be the reference node without loss of generality.

To verify the proposed algorithms, we consider two experimental setups (a) a fixed network of asynchronous nodes and (b) a mobile network of asynchronous nodes. Furthermore, both setups are evaluated for (1) varying number of pairwise communications $K$ for fixed noise on the time markers with standard deviation $\sigma= 10^{-8}$ seconds and (2) varying $\sigma$ in the range $[-90, -70]$ dB seconds for $K=20$. The timing error of $\sigma= 10^{-8}$ seconds (and noise range $[-90, -70]$ dB) translates to a ranging error of $\approx 3.3$ meters (and $\approx[0.33, 33.33]$ meters) for a static network model, since $\text{var}(\tau_{ij}) = c^2\ \times \text{var}(d_{ij})$ with $c= 3 \times 10^{8}$ m/s. Although such high SNR is not usually considered in clock synchronization literature \cite{wu11}, it is typical to achieve meter level accuracies for localization \cite{patwari05,patwari2003}.

\begin{figure*} \centering
  \begin{subfigure}[b]{0.32\textwidth}
    \centering
    \includegraphics[scale=0.35]{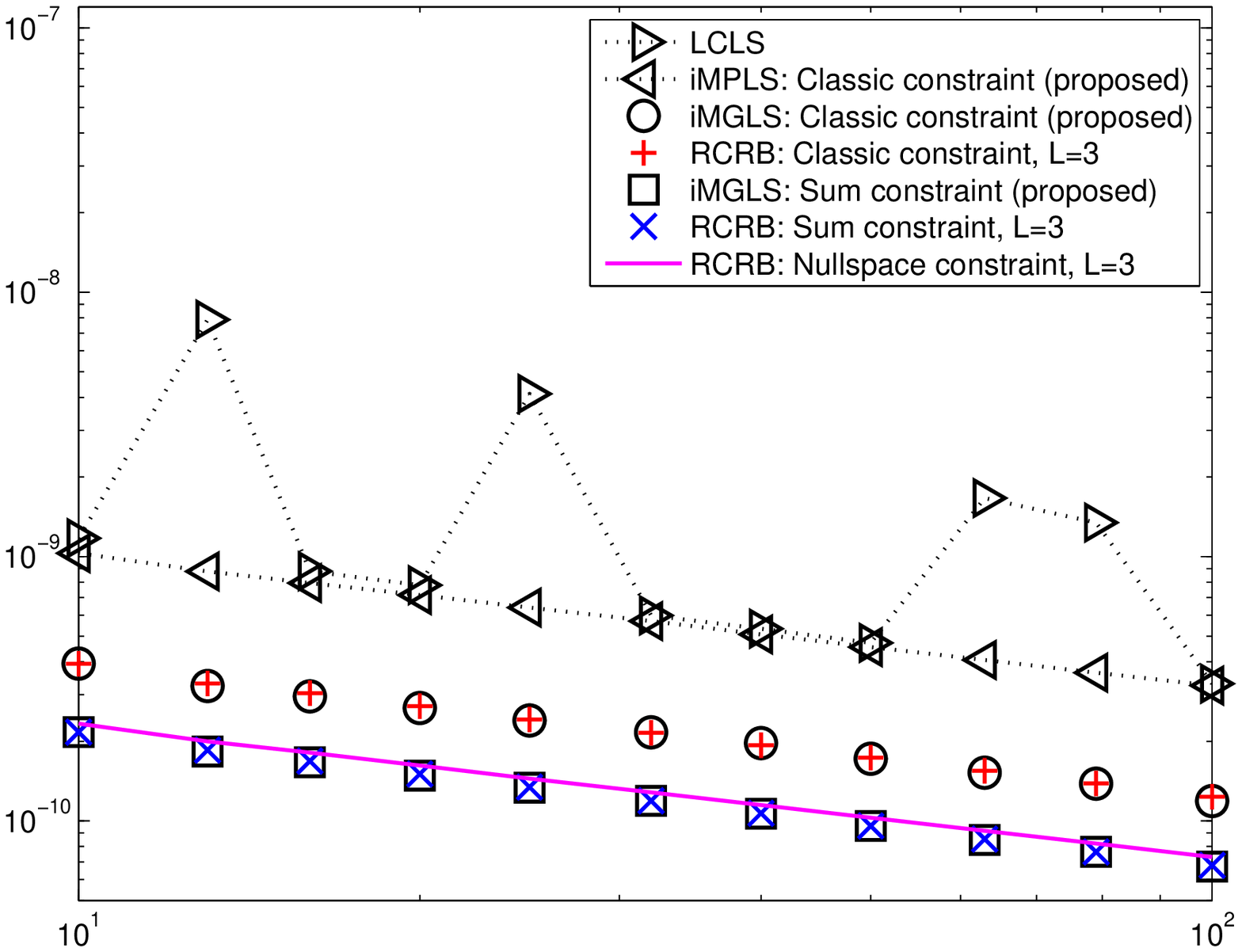}
    \rput(0.5,+.25){\tiny{Number of two-way communications (K)}}
    \rput(0.4, 5.2){\small{RMSE of clock skew ($\hat{\bomega}$)}}
    \caption{}
  \end{subfigure}
  \begin{subfigure}[b]{0.32\textwidth}
    \centering
    \includegraphics[scale=0.35]{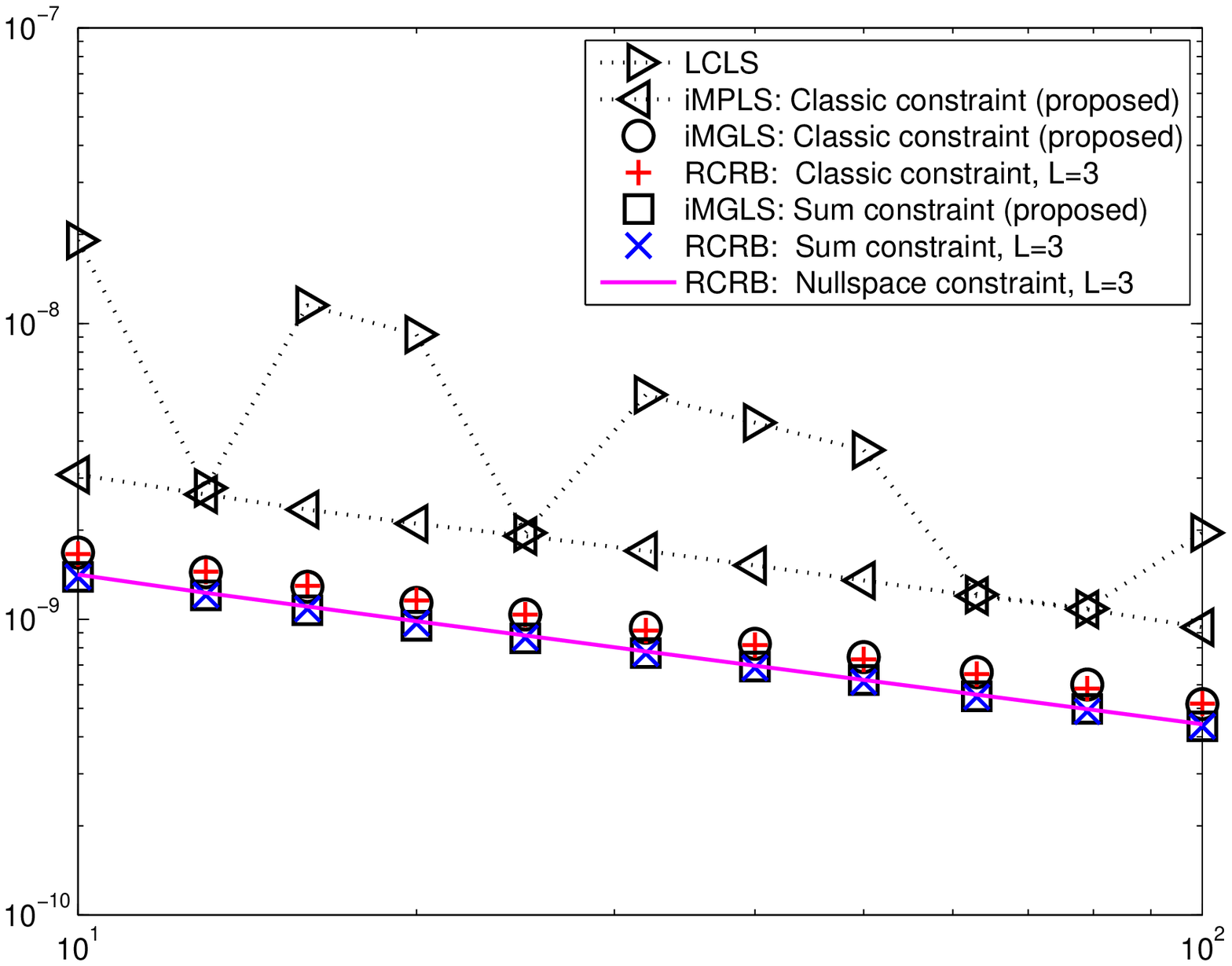}
    \rput(0.5,+.25){\tiny{Number of two-way communications (K)}}
    \rput(0.4, 5.2){\small{RMSE of clock offset ($\hat{\bphi}$)}}
    \caption{}
  \end{subfigure}
  \begin{subfigure}[b]{0.32\textwidth}
    \centering
    \includegraphics[scale=0.35]{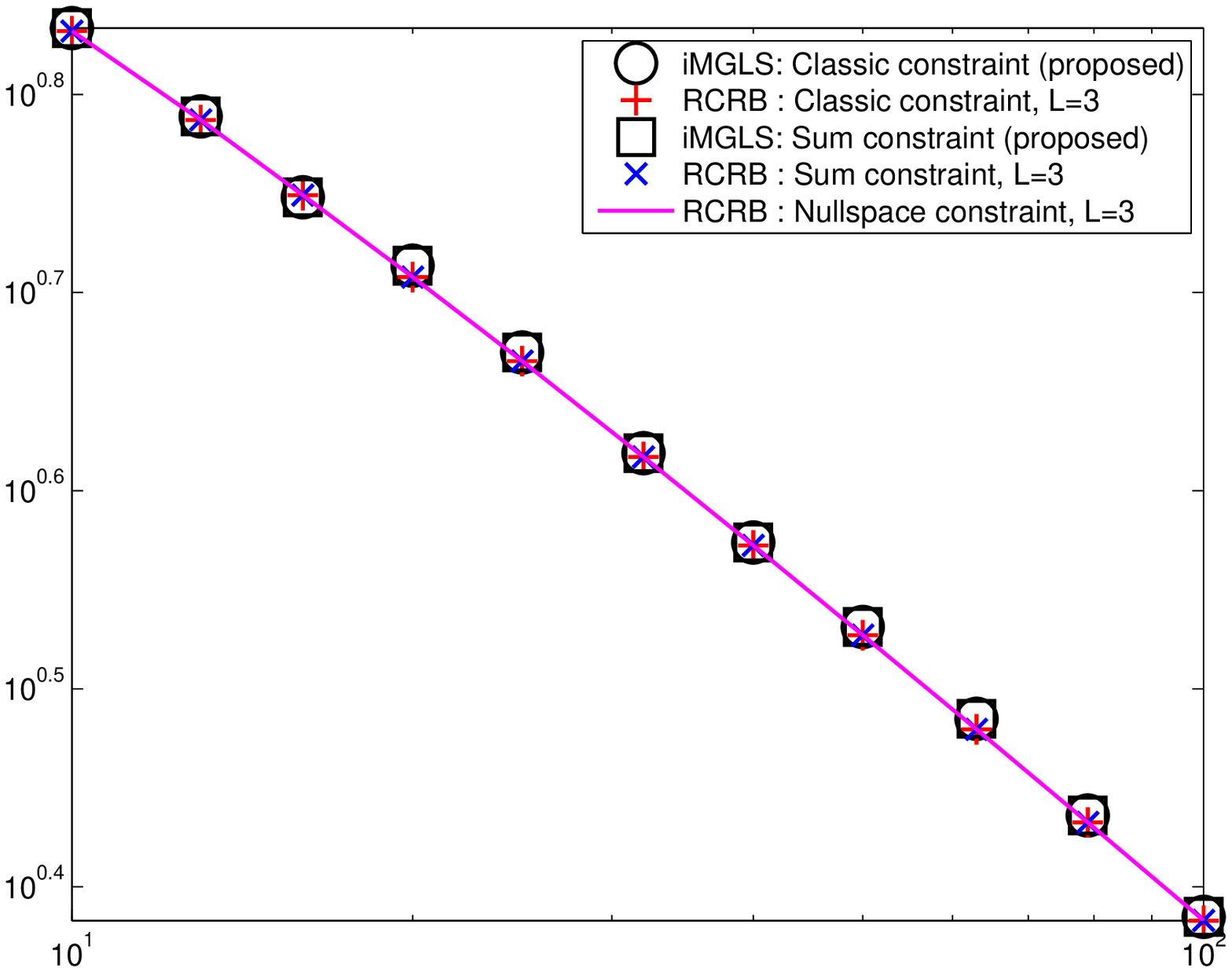}
    \rput(0.5,+.25){\tiny{Number of two-way communications (K)}}
    \rput(0.4, 5.2){\small{RMSE of distance ($\hat{\bd}$)}}
    \caption{}
  \end{subfigure}
  \caption{\small \emph{\textbf{Mobile network: Varying $K$:}} RMSEs (and RCRBs) of (a) clock skew, (b) clock offset and (c) distances for varying number of communications ($K$) between the $N=10$ \emph{mobile} nodes for $\sigma=10^{-8}$ seconds}
  \label{fig:mobileK}
\end{figure*}


\begin{figure*} \centering
  \begin{subfigure}[b]{0.32\textwidth}
    \centering
    \includegraphics[scale=0.35]{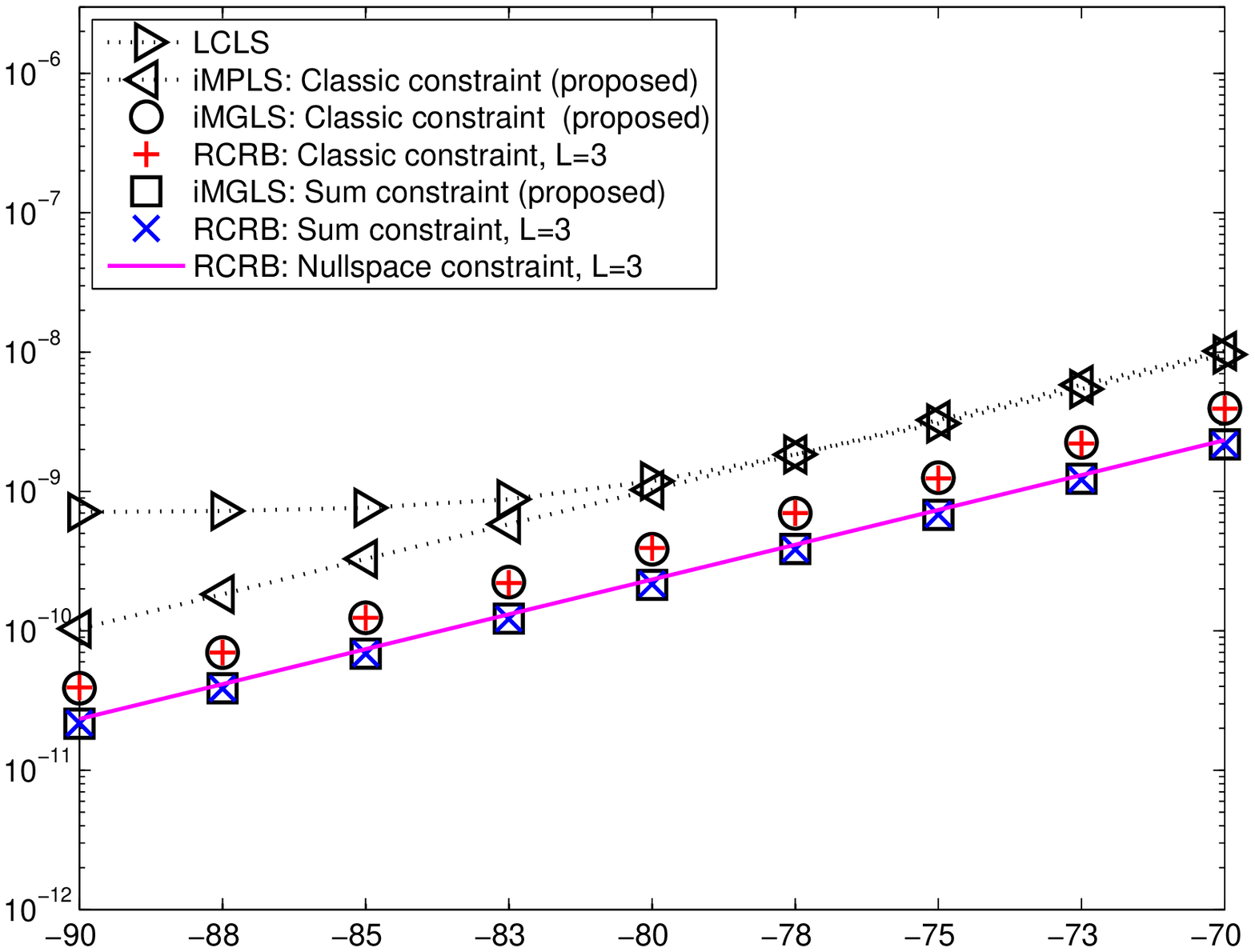}
    \rput(0.5,+.25){\tiny{$10\log_{10}(\sigma)$ [dB second]}}
    \rput(0.4, 5.2){\small{RMSE of clock skew ($\hat{\bomega}$)}}
    \caption{}
  \end{subfigure}
  \begin{subfigure}[b]{0.32\textwidth}
    \centering
    \includegraphics[scale=0.35]{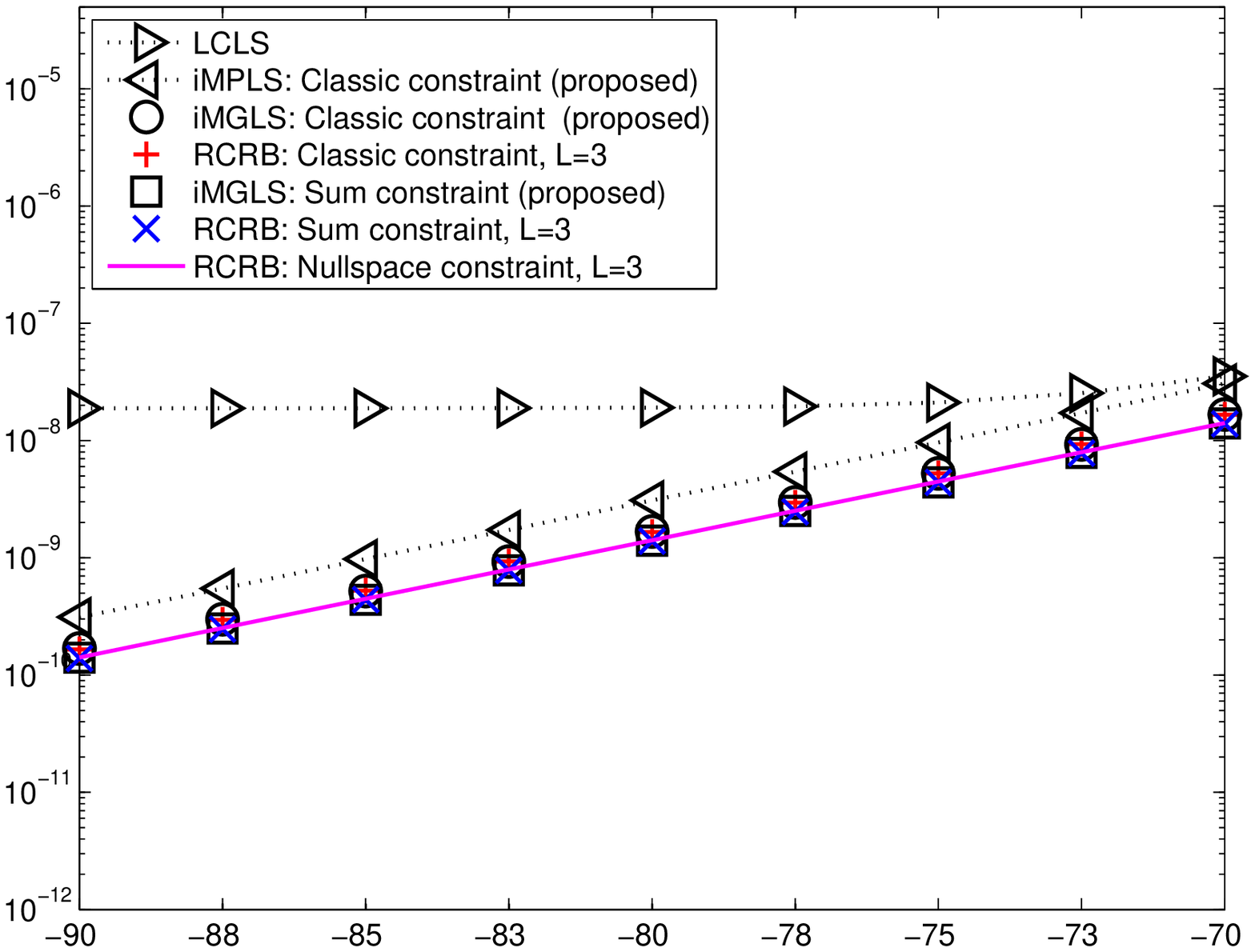}
    \rput(0.5,+.25){\tiny{$10\log_{10}(\sigma)$ [dB second]}}
    \rput(0.4, 5.2){\small{RMSE of clock offset ($\hat{\bphi}$)}}
    \caption{}
  \end{subfigure}
  \begin{subfigure}[b]{0.32\textwidth}
    \centering
    \includegraphics[scale=0.35]{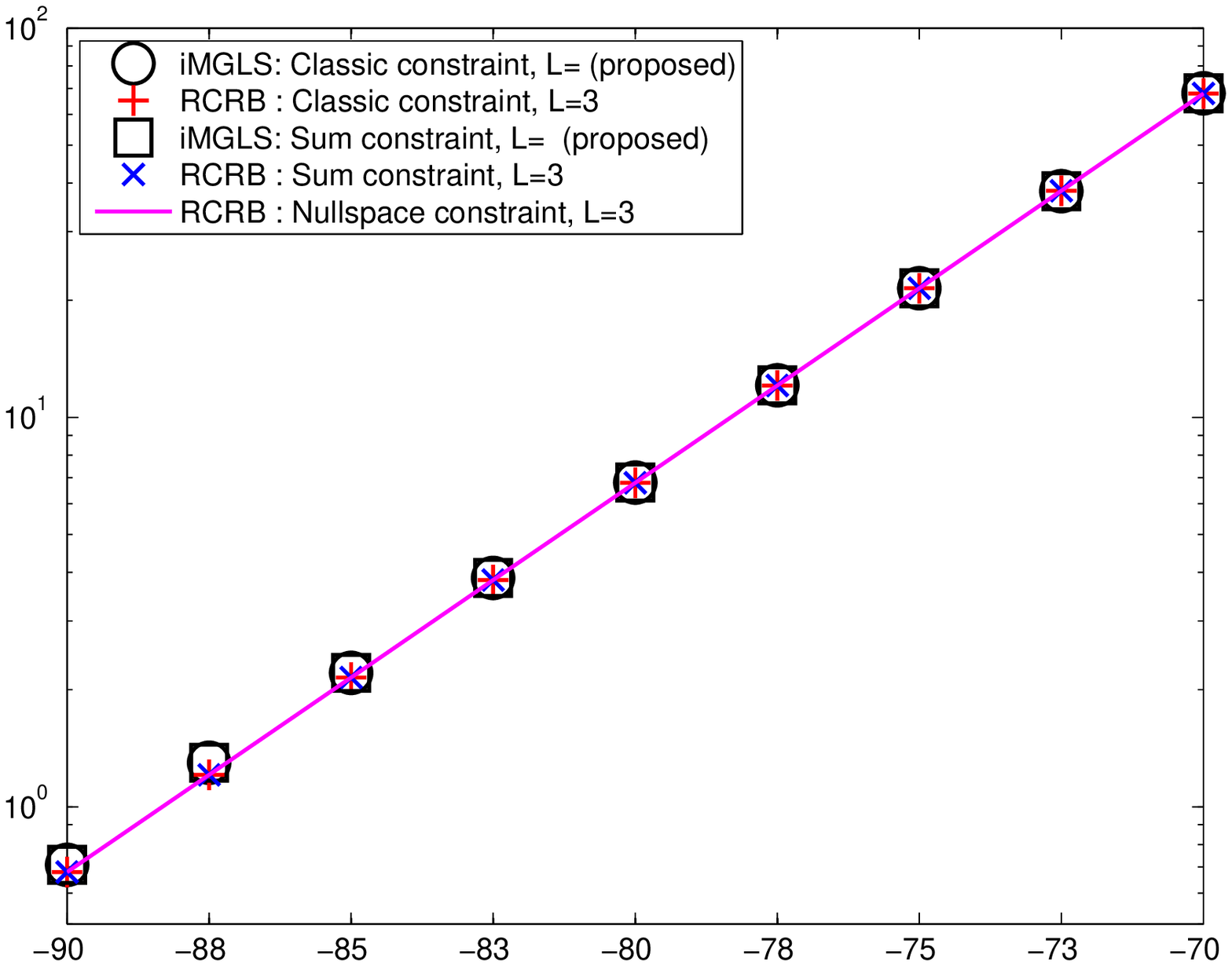}
    \rput(0.5,+.25){\tiny{$10\log_{10}(\sigma)$ [dB second]}}
    \rput(0.4, 5.2){\small{RMSE of distance ($\hat{\bd}$)}}
    \caption{}
  \end{subfigure}
  \caption{\small \emph{\textbf{Mobile network: Varying $\sigma$:}} RMSEs (and RCRBs)  of (a) clock skew, (b) clock offset and (c) distances for a cluster of \emph{mobile} nodes, for varying noise ($\sigma$) on the time measurements with $K=20$.}
  \label{fig:mobileSNR}
\end{figure*}


\subsection{Immobile network} Let the locations of the $N$ nodes be $\bX =\begin{bmatrix} \bx_1, \bx_2, \hdots, \bx_N \end{bmatrix} \in \mathbb{R}^{N \times 2}$ in a $2$ dimensional space, which are arbitrarily chosen to be (\ref{eq:positionValues}), where $\bx_i \in \mathbb{R}^{2 \times 1}$ is the position of the $i$th node. The time invariant propagation delay between the nodes is then \begin{eqnarray} \tau_{ij} &\triangleq& c^{-1}d_{ij}=\ c^{-1}r^{(0)}_{ij}=\      c^{-1}\| \bx_{i} - \bx_{j} \|_2.
\end{eqnarray} The proposed MPLS algorithm (Section \ref{sec:MPLS}) for $L=1$ is independently applied, pairwise from node $1$ to every other node as in \figurename\ \ref{fig:figVariations}a to estimate all the unknown clock skews ($\bomega$), clock offsets ($\bphi$) and range parameters ($\br$). For the entire network, the proposed MGLS (Section \ref{sec:MGLS}) algorithm (with $L=1$) is applied to estimate both the clock parameters $\{\bomega\ \bphi \}$ and the range parameters $\br$. Note that for a fixed network $\bd \triangleq\ \br \in \mathbb{R}^{\barN \times 1}$, where $\bd$ contains all $\barN$ \emph{unique} pairwise distances within the network. \figurename\ \ref{fig:staticK} and \figurename \ref{fig:staticSNR} show the RMSE plots for varying number of communications $K$, for the clock skew $(\bomega)$ and the clock offset $(\bphi)$ and pairwise distances $(\bd)$. The RMSE of clock parameter estimates from the Low Complexity Least Squares (LCLS) solution \cite{leng10} ($L=1$) is also presented for clock skew and offset, which not surprisingly performs similar to the MPLS solution for a fixed network \cite{rajanCAMSAP11}.

The MGLS estimate outperforms the MPLS estimate, which is expected, since the total number of communication channels available for the MGLS estimate is greater than that for MPLS\ \ie $\bar{N} > (N-1)$ for $N\ge2$. Furthermore, the MGLS is shown to achieve the CCRB bounds for $L=1$ for both clock and range parameters since the least square solution is the Minimum Variance Unbiased estimate for the assumed Gaussian noise model. For the given experimental setup, with $10$ns\ ($\approx 3.3 \text{meters})$ noise on the time measurements, distance accuracies improve by an order for $K=100$ two-way communications (Figure \ref{fig:staticK}(c)). Secondly, the nullspace and sum constraints are shown to improve the performance of the clock parameter estimates by about a factor $2$. A discussion on the lower bound of the distance parameter is presented in Appendix \ref{ap:ccrbDistance}. It is worth noting that, the RMSE (and RCRBs) of the clock parameters and distance for the sum constraint is nearly the same as the nullspace constraint.

\begin{figure*} \centering
  \begin{subfigure}[b]{0.32\textwidth}
    \centering
    \includegraphics[scale=0.35]{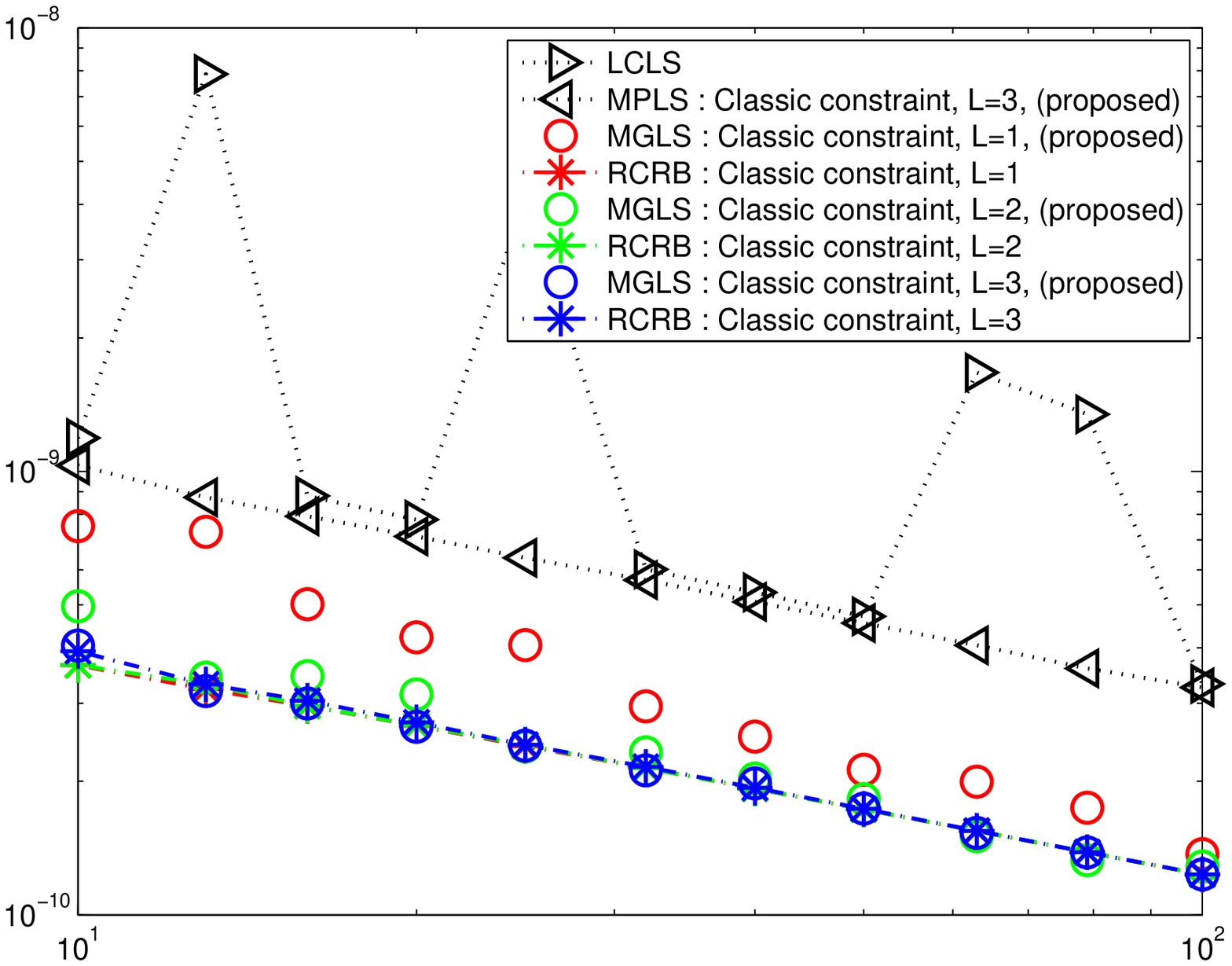}
    \rput(0.5,+.25){\tiny{Number of two-way communications (K)}}
    \rput(0.4, 5.2){\small{RMSE of clock skew ($\hat{\bomega}$)}}
    \caption{}
  \end{subfigure}
  \begin{subfigure}[b]{0.32\textwidth}
    \centering
    \includegraphics[scale=0.35]{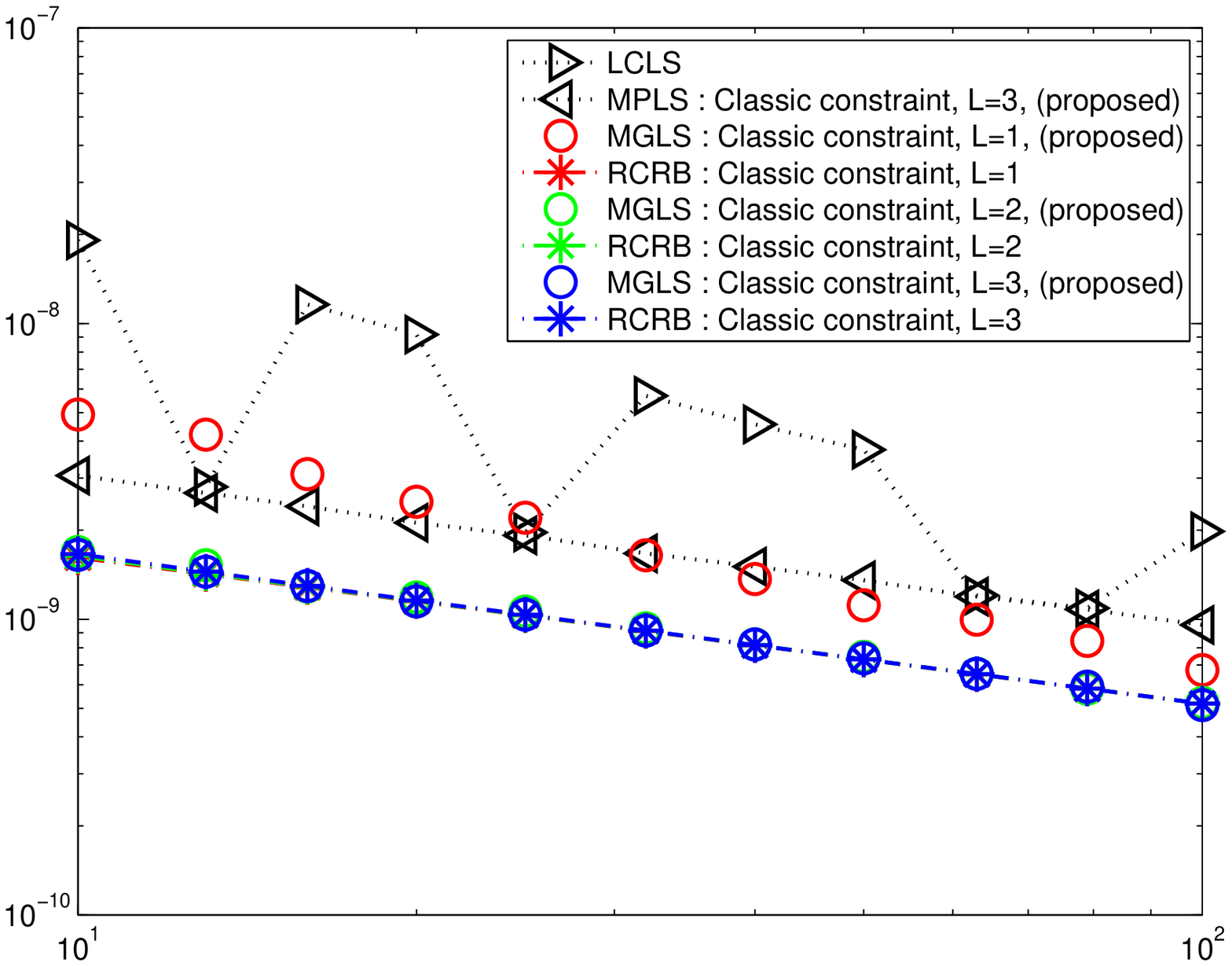}
    \rput(0.5,+.25){\tiny{Number of two-way communications (K)}}
    \rput(0.4, 5.2){\small{RMSE of clock offset ($\hat{\bphi}$)}}
    \caption{}
  \end{subfigure}
  \begin{subfigure}[b]{0.32\textwidth}
    \centering
    \includegraphics[scale=0.35]{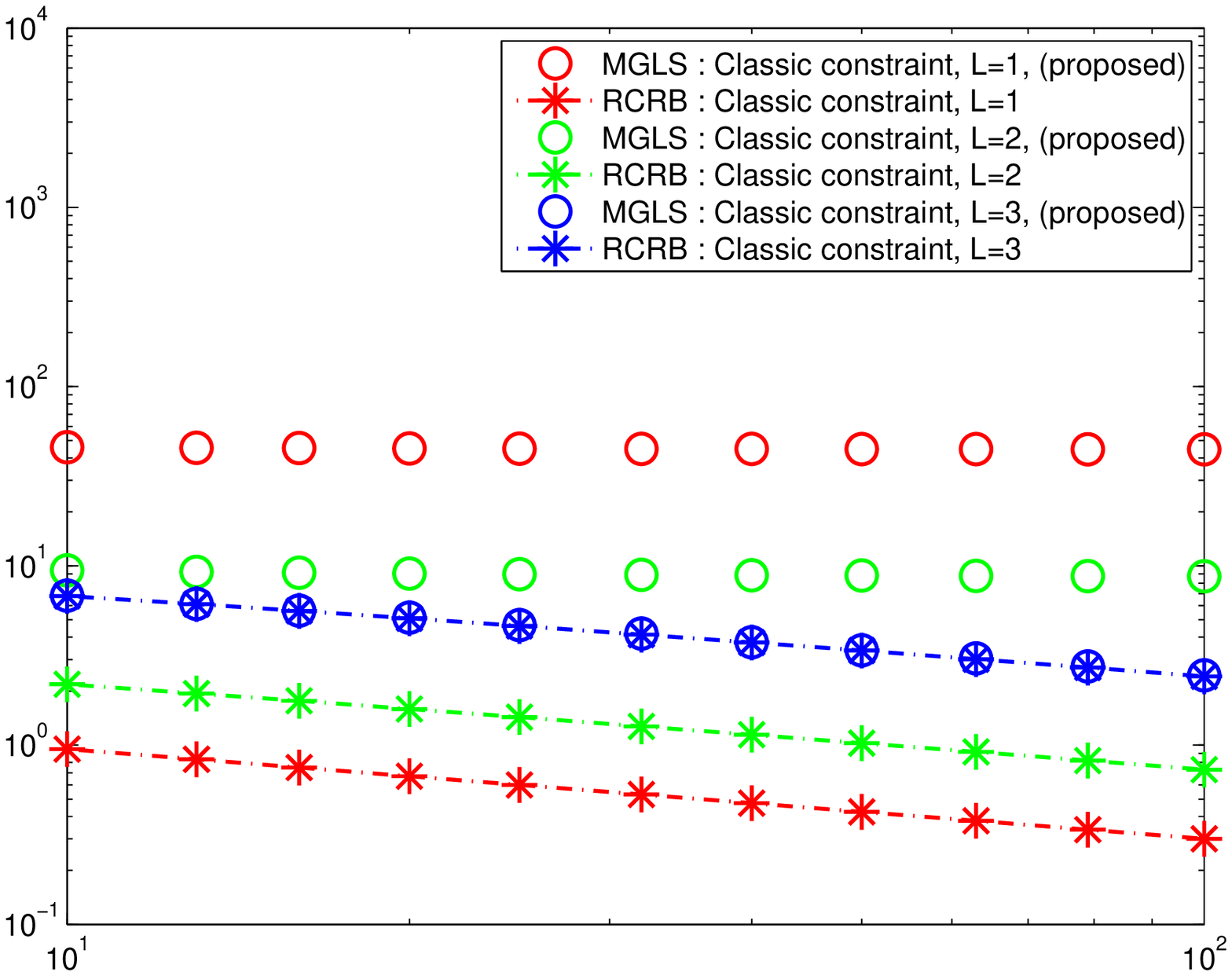}
    \rput(0.5,+.25){\tiny{Number of two-way communications (K)}}
    \rput(0.4, 5.2){\small{RMSE of distance ($\hat{\bd}$)}}
    \caption{}
  \end{subfigure}
  \caption{\small \emph{\textbf{Mobile network: On the choice of $L$ for varying $K$:}} RMSEs (and RCRBs)  of (a) clock skew, (b) clock offset and (c) distances for varying number of communications ($K$) between the $N=10$ \emph{mobile} nodes and different orders of approximation $L$.}
  \label{fig:mobileK_L}
\end{figure*}


\begin{figure*} \centering
  \begin{subfigure}[b]{0.32\textwidth}
    \centering
    \includegraphics[scale=0.35]{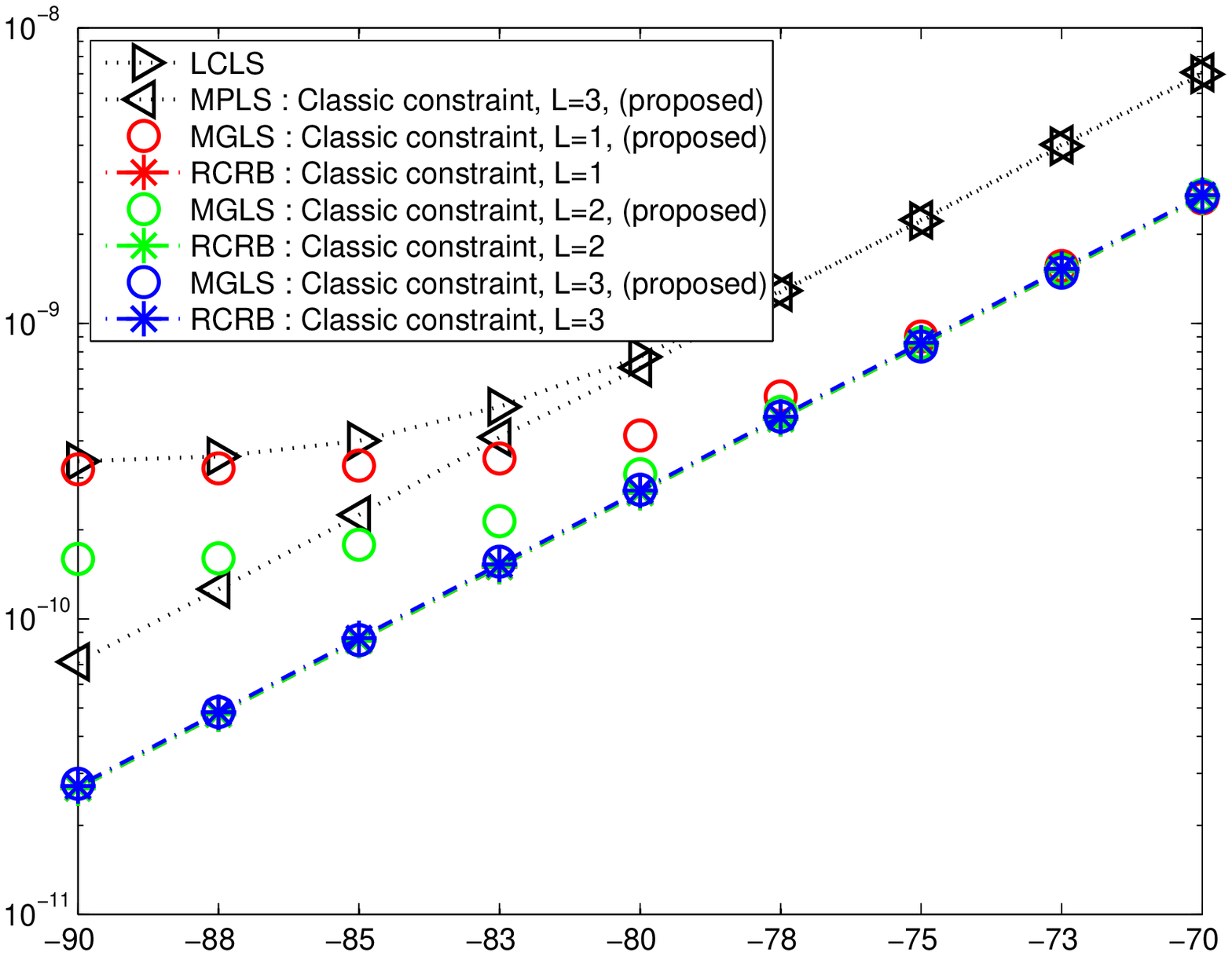}
    \rput(0.5,+.25){\tiny{$10\log_{10}(\sigma)$ [dB second]}}
    \rput(0.4, 5.2){\small{RMSE of clock skew ($\hat{\bomega}$)}}
    \caption{}
  \end{subfigure}
  \begin{subfigure}[b]{0.32\textwidth}
    \centering
    \includegraphics[scale=0.35]{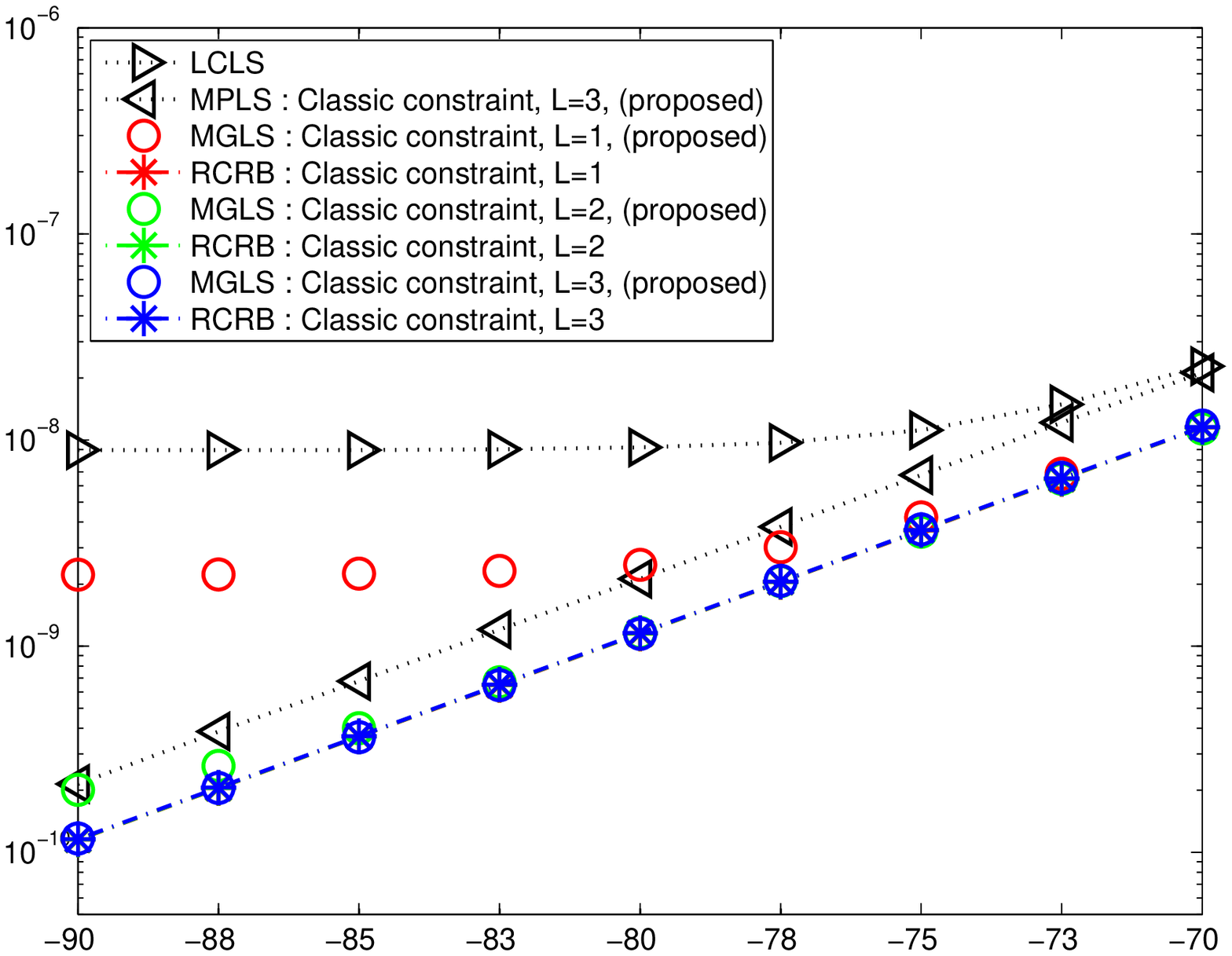}
    \rput(0.5,+.25){\tiny{$10\log_{10}(\sigma)$ [dB second]}}
    \rput(0.4, 5.2){\small{RMSE of clock offset ($\hat{\bphi}$)}}
    \caption{}
  \end{subfigure}
  \begin{subfigure}[b]{0.32\textwidth}
    \centering
    \includegraphics[scale=0.35]{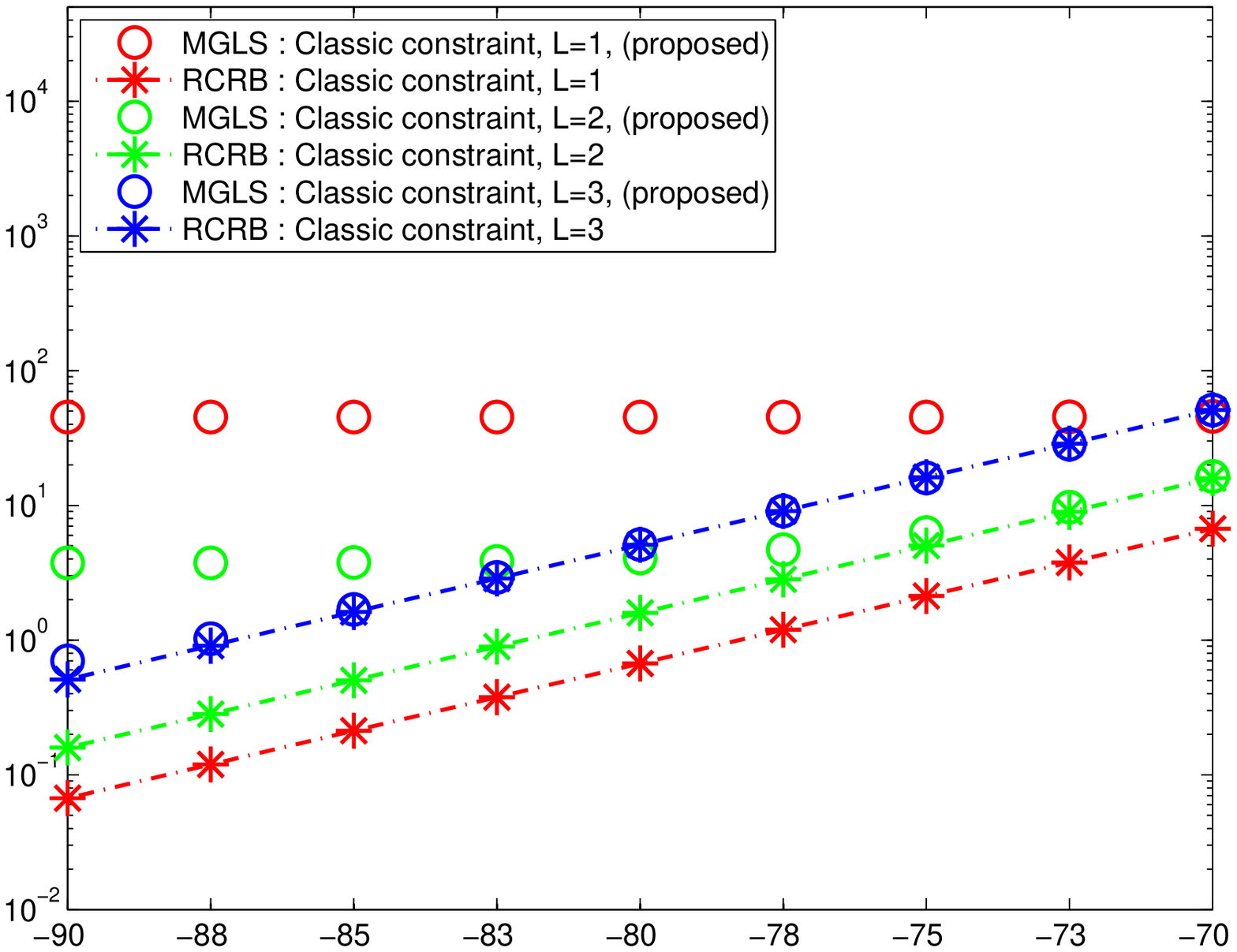}
    \rput(0.5,+.25){\tiny{$10\log_{10}(\sigma)$ [dB second]}}
    \rput(0.4, 5.2){\small{RMSE of distance ($\hat{\bd}$)}}
    \caption{}
  \end{subfigure}
  \caption{\small \emph{\textbf{Mobile network: On the choice of $L$ for varying $\sigma$:}} RMSEs (and RCRBs) (a) clock skew, (b) clock offset and (c) distances, for varying noise ($\sigma$) on the time measurements with $K=20$ and different orders of approximation $L$.}
  \label{fig:mobileSNR_L}
\end{figure*}

\subsection{Mobile network} To investigate the performance of the Least Square solutions for a cluster of mobile nodes, we consider a simple scenario where the nodes are mobile with constant independent velocities. \footnote{Note that the presented model is more general and readily applicable to any motion, as long as $\tau_{ij}(t)$ is a continuous function of time.} The independent constant velocities of the $N$ nodes are given by $\dot{\bX} =\begin{bmatrix} \dot{\bx}_1, \dot{\bx}_2, \hdots, \dot{\bx}_N \end{bmatrix} \in \mathbb{R}^{N \times 2}$ , which similar to the initial positions, are also arbitrarily chosen as (\ref{eq:velocityValues}). Hence, the \emph{true} time varying propagation delay $\tau_{ij}(t)$ \wrt to the clock in node $i$, between the nodes at time instant $k$, is \begin{eqnarray} \label{eq:simulation_dijk}\tau_{ij,k} &\triangleq&  c^{-1}d_{ij,k}=\  c^{-1}\| \tilde{\bx}_{i,k} - \tilde{\bx}_{j,k} \|_2 \end{eqnarray} where \begin{equation}\tilde{\bx}_{i,k} = \bx_i + \dot{\bx}_iT_{ij,k}\quad \forall\ 1 \le i \le N. \end{equation}

\emph{ Note that, even though the nodes are in linear motion, the pairwise distance between the nodes is always non-linear (\ref{eq:simulation_dijk}). In previous cases \cite{rajanCAMSAP11, rajanICASSP12, rajanEUSIPCO13}, fixed range parameters $\br_{ij}= \begin{bmatrix} r^{(0)}_{ij},r^{(1)}_{ij},r^{(2)}_{ij}, \hdots, r^{(L-1)}_{ij} \end{bmatrix}^T \forall i,j \le N$ were used for simulation ensuring the linearity of the joint-time range model, which is unlike the current experimental setup where distance is inherently non-linear and the order of approximation $L$ is unknown.}

Since $L$ is unknown the proposed iMPLS algorithm (Appendix \ref{sec:iMPLS}) is independently applied, pairwise from node $1$ to every other node as in \figurename\ \ref{fig:figVariations}a to estimate all the unknown clock skews, clock offsets and range coefficients. For the given input parameters, the iterative algorithms are observed to converge for $L=3$. For the entire network the iMGLS algorithm (Appendix \ref{sec:iMGLS}) is applied to estimate the clock parameters $[\bomega\ \bphi ]$ and the distances. Observe that unlike the fixed network (with $\barN$ \emph{unique} pairwise distances), the mobile scenario has $\bar{N}K$ \emph{unique} pairwise distances to be estimated, \ie $\bar{N}$ \emph{unique} pairwise distances between the nodes, at all $K$ discrete time instances  during the measurement period $\Delta T$. As before, we investigate the performance of the proposed algorithms for all the $3$ constraints, \ie the classic constraint, nullspace constraint and the sum constraint. All the corresponding RMSEs of the clock skew, offset and distance estimates are plotted in \figurename\ \ref{fig:mobileK} and \figurename\ \ref{fig:mobileSNR} along with their respective RCRB derived in (\ref{eq:crbSigmaEta}) and (\ref{eq:crbDistance}) for various constraints.

The proposed iMPLS algorithm outperforms the LCLS algorithm \cite{leng10} for clock skew and offset estimation of a  mobile network, as shown in \figurename\ \ref{fig:mobileK}. Recall that the LCLS algorithm assumes a fixed network. In addition, numerous outliers are also observed in case of LCLS, since the approximation error of the time-varying distance dominates the gaussian noise under consideration. \emph{Secondly, it is perhaps not surprising that the iMGLS solution achieves the theoretical bounds asymptotically for the clock parameters ($\balpha,\bbeta$) since the linearity of the clock model is ensured via exact parameterization. However, for the non-linear range model in conjunction with the affine clock model, given that the nodes are in independent linear motion (\ref{eq:positionValues}, \ref{eq:velocityValues}), the distance parameters achieving the CCRB at $L=3$ confirms the validity of the joint time-range model.}

In \figurename\ \ref{fig:mobileSNR}, where the RMSE of the proposed algorithms are compared against varying noise variance, the iMPLS  shows considerable improvement over LCLS for high SNR. For lower SNR however, particularly when $\sigma > 1\text{ns} \approx\ 0.3$ meters, the difference between the performances of iMPLS and LCLS is negligible. This is because the noise variance exceeds the magnitude of the velocities (few meters/second in the current experimental setup) and hence, the effect of higher order approximation of the time-varying distance is ineffective.

\subsection{Effect of $L$ on estimation error} \label{sec:sim:EffectOfL} The iterative algorithms (iMPLS, iMGLS) implicity choose the distance approximation order $L$ which minimizes the Least Squares error. To understand the effect of choosing $L$ on the RMSE of the clock and distance parameters, we investigate the performance of MPLS and MGLS algorithms for $L=1,2,3$. \figurename\ \ref{fig:mobileK_L} (varying $K$) and \figurename\ \ref{fig:mobileSNR_L} (varying $\sigma$) show the RMSE and RCRB plots of the proposed algorithm for a single clock reference, \ie the classic constraint.

For the given experimental setup, the RCRBs of the clock parameters are nearly indistinguishable for $L=1,2,3$ (and thus overlay on the plots). However, \figurename\ \ref{fig:mobileK_L}(a) and \ref{fig:mobileK_L}(b) show a factor improvement in the performance of the MGLS algorithm for clock offset and skew. Furthermore, the disparity between $L=1$ and the optimal $L=3$  increases by an order for higher SNR scenarios as presented in \figurename\ \ref{fig:mobileSNR_L}(a) and \ref{fig:mobileSNR_L}(b). A significant advantage of utilizing the proper $L$ is observed in RMSE of the distance parameter in \figurename\ \ref{fig:mobileK_L}(c) and \ref{fig:mobileK_L}(c). As the approximation order increases, the RCRB of the distance (dominated by the Vandermonde-like system) also increases, while the RMSE of the distance estimate steadily decreases with incrementing $L$. An optimality is achieved at $L=3$, when the RMSE of the distance estimate meets the RCRB. Similar to the performance of the clock parameters, for lower SNR the higher order approximation is redundant. Observe in \figurename \ref{fig:mobileSNR_L}, for $\sigma=10^{-7} \text{seconds} \approx 33\ \text{meters}$ with $K=20$, the lower bound and the errors of the distance parameter are equivalent for both $L=1$ and $L=3$, which is not surprising given the velocities are a few meters/second.
\begin{figure} \centering
  \begin{subfigure}[b]{0.5\textwidth}
    \centering
    \includegraphics[scale=0.45]{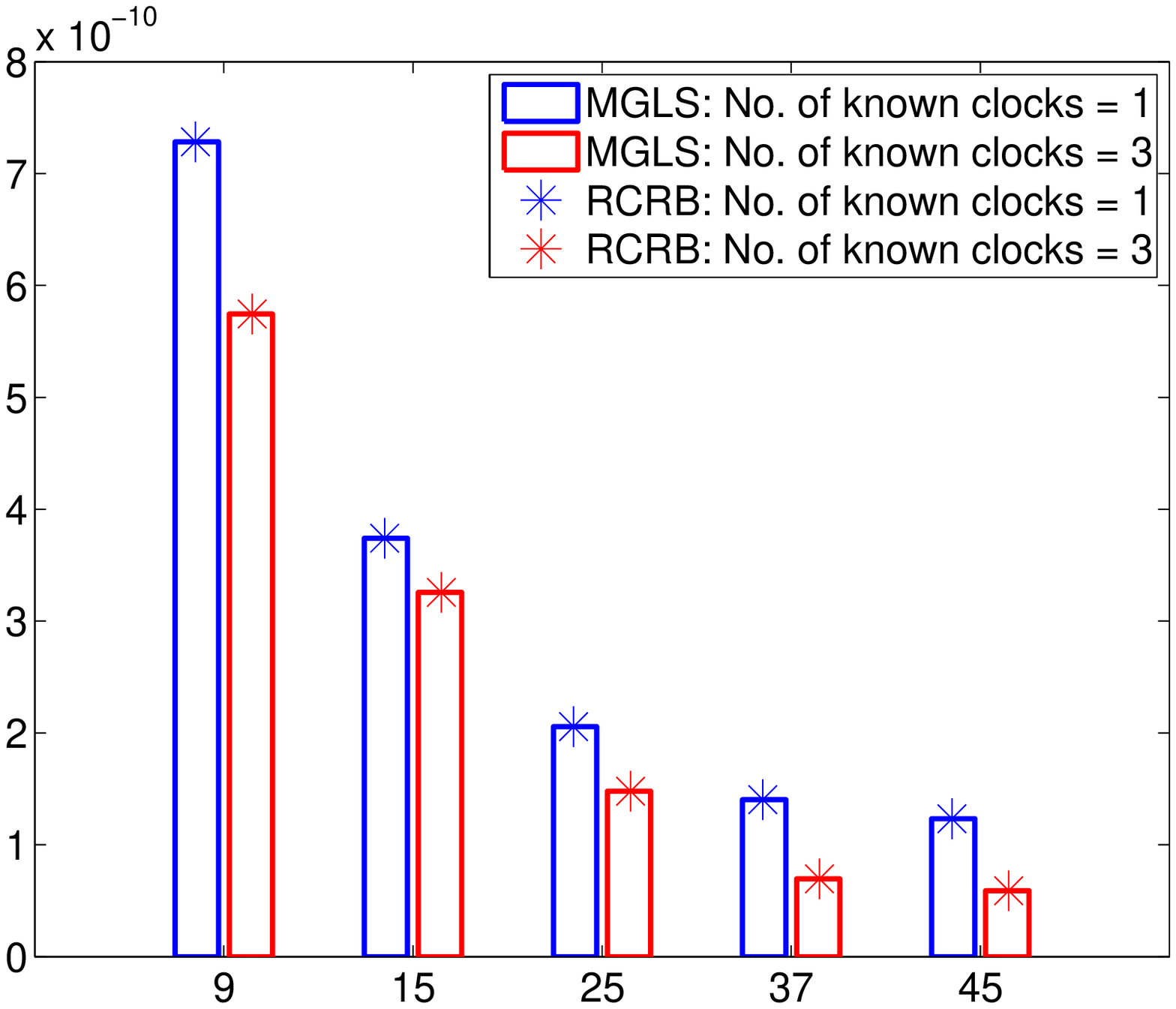}
    \rput(-3.8,-0.1){\small{Number of connected links}}
    \rput*{90}(-8.2, 3.6){\small{RMSE of clock skew ($\hat{\bomega}$)}}
    \caption{}
  \end{subfigure}
  \begin{subfigure}[b]{0.5\textwidth}
    \centering
    \includegraphics[scale=0.45]{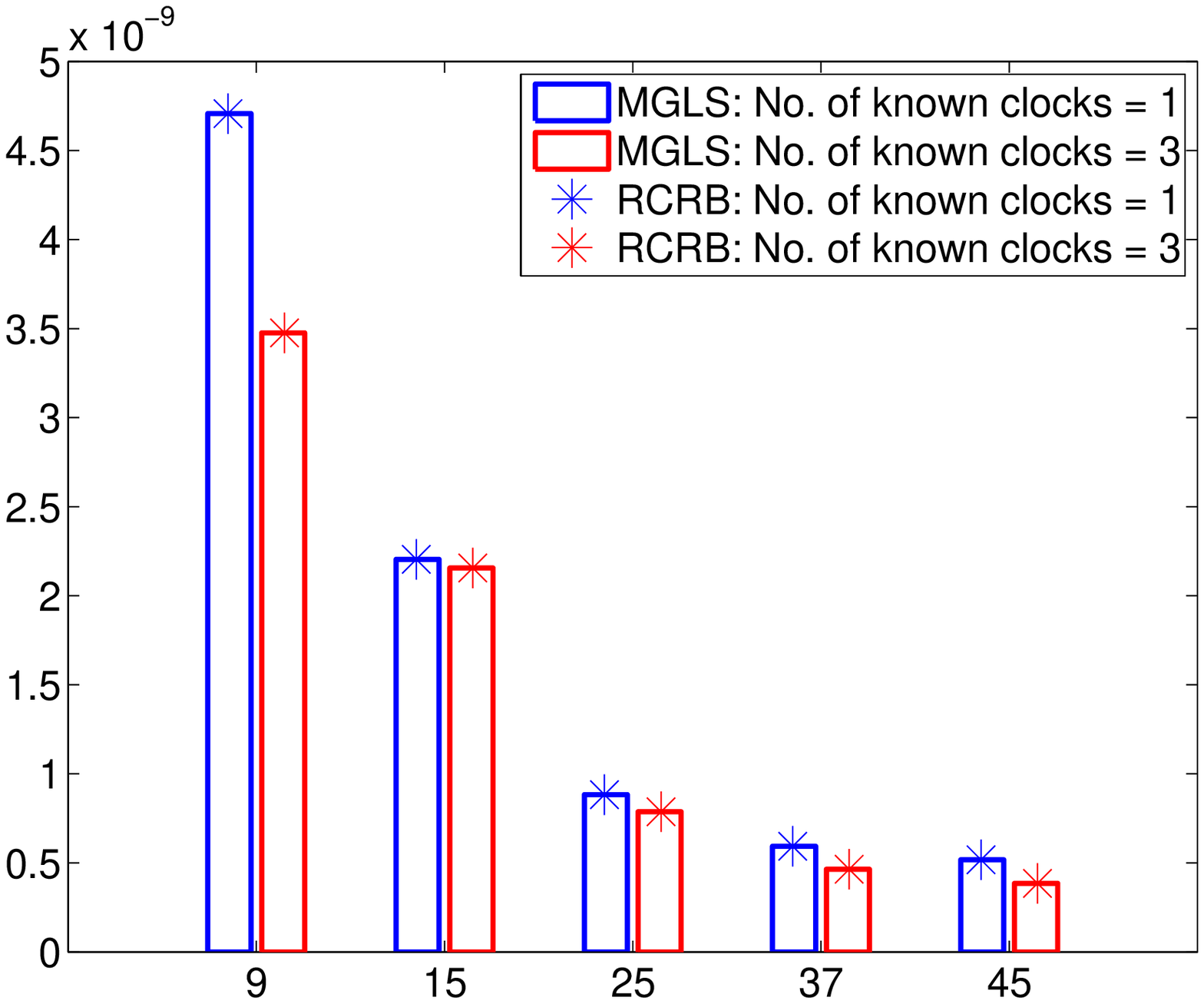}
    \rput(-3.8,-0.1){\small{Number of connected links}}
    \rput*{90}(-8.2, 3.6){\small{RMSE of clock offset ($\hat{\bphi}$)}}
    \caption{}
  \end{subfigure}
  \caption{\small \emph{\textbf{Partially connected Mobile scenario and effect of additional clocks:}} RMSEs (and RCRBs) of (a) clock skew and (b) clock offset for $K=10,\ L=3, \sigma=10$ns, for varying number of connected links.}
  \label{fig:partial}
\end{figure}
\subsection{Extension to partially connected networks}\label{sec:sim:partialNetwork} The proposed MGLS algorithm caters to a full mesh network and can be extended to partial networks for clock synchronization as discussed in \textbf{Remark 3}. For the given mobile network of $N=10$ nodes, the minimum requirement on the number of links is $N-1=9$ (\eg \figurename\ \ref{fig:figVariations}(b)) and for a full mesh network we have $\bar{N}=45$ links. We evaluate the performance of the MGLS algorithm for the synchronization in case of a partially connected network, by varying the number of connected links as $9, 15, 25, 37$ and $45$. The links are arbitrarily chosen such that each node has at least single two-way communication link with one other node in the network, to ensure network wide synchronization. Subsequently, the rows and columns of the  corresponding non-existing links are eliminated from the primary matrix $\bA$ (\ref{eq:dataModelMGLS}). The MGLS algorithm is implemented for $L=3$ with a single clock reference (\ie classic constraint) for $K=10$ and $\sigma=10$ns, and the performance of the clock parameters are presented in \figurename\ \ref{fig:partial}, shown by blue colored markers. Not surprisingly, the RMSE of clock parameters deteriorate with the increase in missing links.

In addition, to emphasize the benefits of the constrained formulation (Section \ref{sec:constrainedBenefits}), we assume that first $3$ clocks of the $10$ node clocks are known in each of the partially connected networks under study. The constraint matrix is then according designed (\eg (\ref{eq:constraintExample_1})) and the performance of the corresponding MGLS solution is presented in \figurename\ \ref{fig:partial}, shown by red colored markers. The incorporation of $2$ additional reference clocks improves the performance of the clock parameters. Furthermore, observe that a partially connected network with $\bar{N}=37$ links and $3$ reference clocks outperforms the full mesh network of $\bar{N}=45$ with a single clock. Such observations can be directly interpreted from the CCRB and the proposed algorithm achieves this CCRB asymptotically, catering readily to such partially connected networks with (or without) apriori information.

\subsection{Summary} We validate the joint-time range model by simulating a two-way time stamp exchange framework for an \emph{asynchronous cluster of mobile nodes}, where the pairwise distances are time varying, and the approximation order of distance $L$ is unknown. The proposed MPLS and MGLS algorithms clearly outperform the prevalent solutions when the nodes are in motion, and in particular for relatively higher SNR on the time markers. More significantly, the variance of the estimated clock parameters and distance achieve the derived CCRB asymptotically. The proposed sum constraint shows an improvement of about factor $2$ in contrast to the classic constraint, and is nearly identical to the performance of the ``optimal'' nullspace constraint, for both mobile and immobile networks. Furthermore, the extension of the proposed algorithms to a partially connected network is simulated for various number of missing links. In addition, the benefits of the constrained framework is shown by studying the effect of multiple clocks in partially connected networks.

\section{Conclusion} The fundamental challenge  has been to jointly estimate clock discrepancies and the time varying distances between a cluster of asynchronous mobile nodes, which is addressed by proposing a novel joint time-range basis. The clock parameters are modeled up to the first order (clock skews, clock offsets) and the pairwise distances between up to a $L-1$th order monomial of \emph{true} time consisting of $L$ range coefficients for each pairwise link. An elegant linear transformation decouples the clock errors from the estimated range parameters. This joint time-range basis has been applied to the proposed generalized TWR scenario and shown to be a linear system of unknown clock and range parameters. (More generally, the joint basis can be applied to other two-way communication frameworks as well.) Subsequently a global least squares solution (MGLS) is proposed, which is in turn an extension of the corresponding distributed pairwise algorithm (MPLS), to estimate all the clock parameters and the pairwise distances at discrete time intervals. Furthermore, when the order $L$ of range approximation is unknown, iterative solutions (iMGLS, iMPLS) are proposed to estimate the apt approximation order for the distance measurement. A novel Constrained \Cramer\ Rao Bound is derived for the presented model and the proposed solutions meet this lower bound asymptotically, which is corroborated by the simulations. As an alternative to the classical single clock reference constraint, we propose the sum constraint and the nullspace constraint which begets a lower variance for clock parameters.

The generalized constrained framework enables users to add more constraints if there is additional information available on the clock and range parameters from other systems, which not surprisingly would increase overall estimation performance. The proposed framework was a full mesh network with two-way communication capability, however a robust synchronization is still feasible despite missing links, including one-way communication. More generally, it can be easily extended to sender-receiver, receiver-receiver, pairwise listening, broadcasting and other prevalent communication schemes (see \cite{wu11} and references therein).

The presented solutions are suited for autonomous networks with minimal a priori knowledge, where the clock and range parameters need to be estimated at \emph{cold start}. Given the pairwise distances, the relative node positions of an anchorless network at every time instant can be estimated using Multi-Dimensional Scaling. In practice, over longer durations, a Kalman filter \cite{Kay1993} can be applied sequentially to track these network parameters, which yields more efficient and optimal estimates with time. The estimated range parameters are viewed merely as coefficients to fit the pairwise distances between the nodes and further investigation on their interpretation is beyond the scope of this article and will be addressed in the followup work \cite{rajan13_camsap}\cite{rajanVelocity}. Finally, although the proposed model is targeted towards anchorless networks, it is readily applicable to anchored scenarios of time, distance and position.

\appendices
\section{Range translation matrix $\bG$} \label{ap:G}
To find an expression for $\bG$, we begin by considering the classic case of a static network of immobile nodes \ie $L=1$. This is a special case of the dynamic range model in (\ref{eq:rangeDefinition}), which has been investigated extensively \cite{rajanCAMSAP11} \cite{leng10}\cite{freris10}. When the nodes are fixed, the propagation delay $\tau_{ij}(t) \triangleq c^{-1}\cR_{ij}(t) = c^{-1}r^{(0)}_{ij}$ is invariant with the \emph{true} time $t$ and following immediately we have \begin{eqnarray} \label{L0_gammaDefinition}
  r^{(0)}_{ij}   &\triangleq& c\gamma^{(0)}_{ij}
\end{eqnarray}A step further, in case of a mobile network, a first order range model is proposed in \cite{rajanICASSP12}, where the translated range model (\ref{eq:rangeTranslation}) for $L=2$ is given by \begin{eqnarray} \label{eq:rangeTranslation_L1}
  \cG_{ij}(t_i) &=& \gamma^{(0)}_{ij} + \gamma^{(1)}_{ij}t_i.
\end{eqnarray} Substituting the equation of ideal \emph{true} time from (\ref{eq:affineTime}) in (\ref{eq:rangeTranslation_L1}), the translated range coefficients in terms of $\alpha_i, \beta_j$ and $r^{(0)}_{ij}$ are \begin{subequations}\label{L1_gammaDefinition} \begin{eqnarray}
    \gamma^{(1)}_{ij}   &\triangleq&  c^{-1}\Big(\alpha_i\ r^{(1)}_{ij}\Big) \\
    \gamma^{(0)}_{ij}   &\triangleq&  c^{-1}\Big(r^{(0)}_{ij} +r^{(1)}_{ij} \beta_i\Big)
  \end{eqnarray}
\end{subequations} and rearranging the terms, \begin{subequations}\label{L1_gammaDefinition} \begin{eqnarray}
    r^{(1)}_{ij}   &\triangleq&  c\ \Big(\alpha^{-1}_i\ \gamma^{(1)}_{ij}\Big) \\
    r^{(0)}_{ij}   &\triangleq&  c\ \Big(\gamma^{(0)}_{ij} -\alpha^{-1}_i \beta_i \gamma_{ij}\Big).
  \end{eqnarray}
\end{subequations} Along similar lines, extending the affine range model to a second order model \cite{rajanEUSIPCO13}  (\ie $L=2$),
we have \begin{eqnarray} \label{eq:rangeTranslation_L2} \cG_{ij}(t_i) &=& \gamma^{(0)}_{ij} + \gamma^{(1)}_{ij}t_i + \gamma^{(2)}_{ij}t^2_i
\end{eqnarray} where an expression for $ \bgamma_{ij}=  [\gamma^{(0)}_{ij}\; \gamma^{(1)}_{ij}\; \gamma^{(2)}_{ij}]$ in terms of the \emph{true} range parameters $\br_{ij}$ and clock errors is obtained by substituting for ideal \emph{true} time from (\ref{eq:affineTime}) in (\ref{eq:rangeTranslation_L2}), which yields \begin{subequations}\label{L2_gammaDefinition}
  \begin{eqnarray}
    \gamma^{(2)}_{ij}   &\triangleq& c^{-1}\Big(\alpha_i^2\ r^{(2)}_{ij}\Big) \\
    \gamma^{(1)}_{ij}   &\triangleq& c^{-1}\Big(\alpha_i r^{(1)}_{ij} + 2\alpha_i\beta_ir^{(2)}_{ij}\Big) \\
    \gamma^{(0)}_{ij}   &\triangleq& c^{-1}\Big(r^{(0)}_{ij} + \beta_ir^{(1)}_{ij} +\beta^2_i r^{(2)}_{ij}\Big)
  \end{eqnarray}
\end{subequations} or alternatively \begin{subequations}\label{L2_rangeDefinition}
  \begin{eqnarray}
    r^{(2)}_{ij}        &\triangleq& c\ \Big(\alpha^{-2}_i\gamma^{(2)}_{ij}\Big) \\
    r^{(1)}_{ij}        &\triangleq& c\ \Big(\alpha^{-1}_i\gamma^{(1)}_{ij} -2\alpha^{-2}_i\beta_i\gamma^{(2)}_{ij}\Big) \\
    r^{(0)}_{ij}        &\triangleq& c\ \Big(\gamma^{(0)}_{ij} + \alpha^{-1}_i\beta_i\gamma^{(1)}_{ij} -\alpha^{-2}_i\beta^2_i \gamma^{(2)}_{ij}\Big)
  \end{eqnarray}
\end{subequations} More generally, for any $L \ge 1$, the $l$th order translated range coefficient $\bgamma^{(l)}_{ij}$ for the node pair $\{i,j\}$ is by symmetry \begin{equation}
  r^{(l)}_{ij}   \triangleq
  c\sum^{L-1}_{\bar{l}=l} \begin{pmatrix} \bar{l} \\ l \end{pmatrix} \alpha_i^{- \bar{l}}  (-\beta_i)^{\bar{l}-l}\gamma^{(\bar{l})}_{ij} \quad \forall\ l=0,1,\hdots, L
  \label{eq:rangeijCrudeDefinition}
\end{equation} which for the sake of notational brevity can be written as \begin{eqnarray}
  \label{eq:rangeBasisPairwise}
  \br_{ij}  =\  \tilde{\bG}_i \bgamma_{ij} &\Leftrightarrow& \bgamma_{ij}  = \tilde{\bG}^{-1}_i \br_{ij}
\end{eqnarray} where $\tilde{\bG}_i \in \mathbb{R}^{L \times L}$, $\forall\ l=0,1,\hdots L \text{and}\ \bar{l}=l+1,l+2,\hdots L$  \begin{eqnarray} \label{eq:rangeijVectorDefinition}
\{ \tilde{\bG}_i \}_{l+1, \bar{l}+1}=&   c \begin{pmatrix} \bar{l} \\ l \end{pmatrix} \alpha_i^{-\bar{l}}  (-\beta_i)^{\bar{l}-l}
\end{eqnarray} is a triangular matrix contains the clock discrepancies of node $i$.

For the entire network of $\barN$ \emph{unique} pairwise links, we have \begin{eqnarray} \label{eq:tilderangeBasis}  \tilde{\br} = \tilde{\bG} \tilde{\bgamma} &\Leftrightarrow& \tilde{\bgamma} = \tilde{\bG}^{-1} \tilde{\br}
\end{eqnarray} where $\tilde{\bgamma} = \text{vec}(\bGamma^{T})= \bP^{T}\bgamma$  and $\tilde{\br} = \text{vec}(\bR^{T})= \bP^{T}\br$, where $\bP \in \mathbb{R}^{\barN L \times \barN L}$ is a permutation matrix. The transformation matrix $\tilde{\bG} \in \mathbb{R}^{\bar{N}L \times \bar{N}L}$ is given by \begin{eqnarray} \label{eq:tildeRangeGammaTranslation}
\tilde{\bG}=  \text{blkdiag}\big(\bI_{N-1}\otimes\tilde{\bG}_1,\ \bI_{N-2}\otimes\tilde{\bG}_2,\ \hdots, \bI\otimes\tilde{\bG}_{N-1} \big)
\end{eqnarray} which is only dependent on the clock calibration parameters $\{\balpha, \bbeta\}$ of the network. Finally, defining \begin{equation}
\label{eq:rangeGammaTranslation}
\bG= \bP\tilde{\bG}\bP^T
\end{equation} we have \begin{eqnarray} \label{eq:rangeBasis_copy}
\br = \bG\bgamma &\Leftrightarrow& \bgamma = \bG^{-1} \br,
\end{eqnarray} which gives us a unique relation between the \emph{true} range parameters and the translated range parameters, in the presence of clock errors. It is evident from (\ref{eq:rangeBasis}) that the range parameters can be extracted uniquely from the modified range parameters despite clock discrepancies, provided $\bG$ \ie the clock calibration parameters $\begin{bmatrix} \balpha& \bbeta \end{bmatrix}$ are known. Furthermore, in the absence of clock errors, \ie $\balpha=\b1_N$ and $\bbeta=\bzero_N$, then $\bG=c\bI_{\bar{N}L}$ and following immediately $\br= c\bgamma$. Observe that, for a given node pair $(i,j)$ although the translated parameters ($\bgamma_{ij}$) are dependent on the choice of clock reference $i$ or $j$, the true range parameters $(\br_{ij})$ remain unique to a given node pair.

\section{iterative Mobile Pairwise Least Squares (iMPLS)}\label{sec:iMPLS} For a given distance approximation order $l$, the pairwise cost function (\ref{eq:mplsSolution}) can be rewritten as \begin{eqnarray}
\Hat{\btheta}_{ij,l}  \label{eq:mplsSolution_l}
=   \arg \min_{\btheta_{ij,l}} \; \epsilon_{ij,l}= (\bA_{ij,l}^T\bA_{ij,l})^{-1}\bA_{ij,l}\bb_{ij}
\end{eqnarray} where \begin{eqnarray}
\label{eq:mplsError}
\epsilon_{ij,l} &=& (\bA_{ji,l}\btheta_{ij,l} - \bb_{ij})^T(\bA_{ji,l}\btheta_{ij,l} - \bb_{ij}) \\
\label{eq:AijDefinition_l}
\bA_{ji}  &=&               [-\bt_{ji} \quad -\b1_{K} \quad  \bE_{ij}\bV_{ij,l}],  \end{eqnarray}, $\bV_{ij,l}=\begin{bmatrix} \bt^{\odot 0}_{ij}  & \bt^{\odot 1}_{ij} & \hdots, & \bt^{\odot l-1}_{ij} \end{bmatrix}$, $\btheta_{ij,l}= [\alpha_j \quad \beta_j \quad \bgamma^T_{ij}]^T$ and $\bb_{ij}=-\bt_{ij}$. More generally, when $l$ is unknown, we briefly describe the iterative Mobile Pairwise Least Squares (iMPLS) algorithm for a pair of nodes, using the well known order recursive least squares \cite{Kay1993}. \begin{algorithm} \caption{iterative Mobile Pairwise Least Squares (iMPLS)}
\label{alg:MPLS}
\begin{algorithmic}
  \STATE{\bf Initialize:}
  \STATE 1) For $l=0$ define $\bA_{ji,l} \triangleq \bA_{ji,0}$ from (\ref{eq:AijDefinition_l})
  \STATE 2) Define $\bar{\bA}_{ji,l} \triangleq\ \bar{\bA}_{ji,0}  = (\bA_{ji,0}^T\bA_{ji,0})^{-1}$
  \STATE 3) Estimate $\hat{\btheta}_{j,l} \triangleq \hat{\btheta}_{j,0}$ using (\ref{eq:mplsSolution_l})
  \STATE 4) Estimate LSE $\epsilon_{ij,l} \triangleq\ \epsilon_{ij,0}$ from (\ref{eq:mplsError})
  \STATE 3) Define $m=3$ and $\delta_0 = \epsilon_{ij,0}/m$
    \WHILE{$\delta_l >\ \delta_{ij,tol}$}
    \STATE 4) Update inverse $\bar{\bA}_{ij,l+1}$ using (\ref{eq:updateMPLSInverse})
    \STATE 5) Estimate $\hat{\btheta}_{ij,l+1}$ from (\ref{eq:updateMPLSTheta})
    \STATE 6) Update least squares error $\epsilon_{ij,l+1}$ using (\ref{eq:mplsSolution_l})
    \STATE 7) Update $\delta_l \leftarrow\ (\epsilon_{ij,l+1}-\epsilon_{ij,l})/(m+1)$
    \STATE 8) Update $l \leftarrow\ l+1$, $m \leftarrow\ m+1$, $\epsilon_{ij,l} \leftarrow\ \epsilon_{ij,l+1}$
  \ENDWHILE
\end{algorithmic}
\end{algorithm}

\begin{figure*}
\normalsize
\setcounter{eqnCounter2}{\value{equation}}
\begin{eqnarray}
\label{eq:updateMPLSInverse}
\bar{\bA}_{ji,l+1}
&=& \begin{bmatrix}
\bar{\bA}_{ji,l} + \frac{\bar{\bA}_{ji,l}\bA^T_{ji,l}\ba_{ij,l+1}\ba_{ij,l+1}^T\bA_{ji,l}\bar{\bA}_{ji,l}}{\ba^T_{l+1}\bP^{\perp}_{ij,l}\ba_{l+1}} &
\frac{\bar{\bA}_{ji,l}\bA^T_{ji,l}\ba_{ij,l+1}}{\ba^T_{l+1}\bP^{\perp}_{ij,l}\ba_{l+1}} \vspace{1mm} \\
\frac{\ba^T_{ij,l+1}\bA_{ji,l}\bar{\bA}_{ji,l}}{\ba^T_{l+1}\bP^{\perp}_{ij,l}\ba_{l+1}} &
\frac{1}{\ba^T_{ij,l+1}\bA_{ji,l}\bar{\bA}_{ji,l}}
\end{bmatrix} \\
\label{eq:updateMPLSTheta}
\btheta_{j,l+1} &=&
\begin{bmatrix} \btheta_{j,l} -  \frac{\bar{\bA}_{ji,l}\bA^T_{ji,l}\ba_{ij,l+1}\ba_{ij,l+1}^T\bP^{\perp}_{ij,l}\bb_{ij}}{\ba^T_{l+1}\bP^{\perp}_{ij,l}\ba_{l+1}} \\
\frac{\ba^T_{ij,l+1}\bP^{\perp}_{ij,l}\bb_{ij}}{\ba^T_{ij,l+1}\bP^{\perp}_{ij,l}\ba_{ij,l+1}}
\end{bmatrix} \\
\label{eq:updateMPLSError}
\epsilon_{ij,l+1} &=&
\epsilon_{ij,l} - \frac{(\ba^T_{ij,l+1}\bP^{\perp}_{ij,l}\bb_{ij})^2}{\ba^T_{ij,l+1}\bP^{\perp}_{ij,l}\ba_{ij,l+1}}
\end{eqnarray}where \begin{equation}
\ba_{ij,l+1} = \be_{ij}\odot\bt^{\odot\ l+1}_{ij} \quad\ \text{and}\ \quad\ \bP^{\perp}_{ij,l}= \bI - \bA_{ji,l}\bar{\bA}_{ji,l}\bA^T_{ji,l} \nonumber
\end{equation}
\setcounter{equation}{\value{eqnCounter2}}
\hrulefill
\vspace*{1pt}
\end{figure*}
\addtocounter{equation}{3}

\section{iterative Mobile Global Least Squares (iMGLS)} \label{sec:iMGLS} Similar to the pairwise model, we propose an iterative Mobile Global Least Squares solution to dynamically estimate all the clock and range parameters for a cluster of mobile nodes, when the range order is unknown. Note that for a given $l$, the KKT solution (\ref{eq:mglsSolution}) is \begin{eqnarray}\label{eq:mglsSolution_l} \hat{\btheta}_l
=   \arg \min_{\btheta_l} \; \epsilon_l= \bB_l^{-1}\underline{\bb} \label{eq:mglsSolution_l}
\end{eqnarray} where \begin{eqnarray}
\label{eq:mglsError_l}
\epsilon_l &=& (\bB_l\btheta_l - \underline{\bb})^T(\bB_l\btheta_l - \underline{\bb}), \\
\label{eq:BDefinition_l}
\bB_l  &=&  \begin{bmatrix} 2\bA_l^T\bA_l & \bC_l^T \\  \bC_l  & \bzero_{N_2, N_2}\\ \end{bmatrix},
\end{eqnarray} $\bA_l= \begin{bmatrix} \bT& \bH& \bE\bV_l \end{bmatrix}$, $\bV= \begin{bmatrix} \bI_{\barN} \otimes \b1_K&  \bar{\bT}^{\odot 1}& \hdots& \bar{\bT}^{\odot L-1} \end{bmatrix}$ and $\underline{\bb}= \begin{bmatrix} \bzero^T& \bb^T \end{bmatrix}^T$

\begin{algorithm}
\caption{iterative Mobile Global Least Squares (iMGLS)}
\label{alg:MGLS}
\begin{algorithmic}
  \STATE{\bf Initialize:}
  \STATE 1) For $l=0$, define $\bB_l \triangleq \bB_0$ using (\ref{eq:BDefinition_l})
  \STATE 2) Estimate $\widehat{\underline{\btheta}}_l \triangleq \widehat{\underline{\btheta}}_0$ using (\ref{eq:mglsSolution_l})
  \STATE 3) Estimate LSE $\epsilon_l \triangleq\ \epsilon_0$ from (\ref{eq:mglsError_l})
  \STATE 4) Define $m=2N + \barN$ and $\delta_l = \epsilon_0/m$.
  \WHILE{$\delta_l>\ \delta_{tol}$}
    \STATE 4) Estimate $\hat{\underline{\btheta}}_{l+1}$ using (\ref{eq:mglsSolution})
    \STATE 5) Obtain least squares error $\epsilon_{l+1}$ using (\ref{eq:mglsError_l})
    \STATE 6) Update $\delta_l \leftarrow\ (\epsilon_{l+1}-\epsilon_{l})/(m+1)$
    \STATE 7) Update $l \leftarrow\ l+1$, $m \leftarrow\ m+\barN$, $\epsilon_l \leftarrow\ \epsilon_{l+1}$
  \ENDWHILE
\end{algorithmic}
\end{algorithm}

\section{Jacobian $\bJ_{\theta\eta}$} \label{sec:appendixJacobian} The Jacobian $\bJ_{\theta\eta}$ in (\ref{eq:crbSigmaEta}) is given by\begin{eqnarray}
\label{eq:JacobianMGLS}
\bJ_{\theta{\eta}}
& \triangleq &
\begin{bmatrix}
\dfrac{\partial \bEta}{\partial\btheta^T}
\end{bmatrix}
\triangleq
\begin{bmatrix}
\dfrac{\partial\bEta}{\partial\balpha^T} \quad
\dfrac{\partial\bEta}{\partial\bbeta^T}  \quad
\dfrac{\partial\bEta}{\partial\bgamma^T}
\end{bmatrix} \nonumber \vspace{1mm} \nonumber \\
&=&
\begin{bmatrix}
\dfrac{\partial\bomega}{\partial\balpha^T} \quad \dfrac{\partial\bomega}{\partial\bbeta^T}\quad \dfrac{\partial\bomega}{\partial\bgamma^T}\\
\dfrac{\partial\bphi}{\partial\balpha^T} \quad   \dfrac{\partial\bphi}{\partial\bbeta^T} \quad  \dfrac{\partial\bphi}{\partial\bgamma^T} \\
\dfrac{\partial\br}{\partial\balpha^T} \quad     \dfrac{\partial\br}{\partial\bbeta^T} \quad    \dfrac{\partial\br}{\partial\bgamma^T}
\end{bmatrix} \nonumber \\
&=&
\begin{bmatrix}
-\cA^2 &              \bzero_{N,N}&       \bzero_{N,\bar{N}L} \\
\cA^2\cB &            -\cA  &             \bzero_{N,\bar{N}L} \\
\dot{\bG}^{\alpha} &  \dot{\bG}^{\beta}&  \bG
\end{bmatrix}
\end{eqnarray} where $\cA= \diag(\balpha)^{-1} \in \mathbb{R}^{N\times N}$, $\cB= \diag(\bbeta) \in \mathbb{R}^{N\times N}$, $\bG$ is the transformation matrix defined in (\ref{eq:rangeGammaTranslation}).

\begin{figure*}\normalsize
\vspace*{1pt} 
\setcounter{eqnCounter4}{\value{equation}}
\begin{eqnarray}
\label{eq:rangeGammaTranslationDerivativeAlpha}
\dot{\bG}^{\balpha}   &=& \dfrac{\partial\bP\bG\bP^T}{\partial\balpha^T}
                      = \bP \begin{pmatrix}\dfrac{\partial\bG}{\partial\balpha^T}\end{pmatrix}\bP^T
                      = \bP \Bigl(\text{blkdiag}\big(\bI_{N-1}\otimes\dot\bG^{\alpha}_1, \bI_{N-2}\otimes\dot\bG^{\alpha}_2, \hdots, \bI\otimes\dot\bG^{\alpha}_{N-1} \big)\Bigr) \bP^T \\
\label{eq:rangeGammaTranslationDerivativeBeta}
\dot{\bG}^{\bbeta}    &=& \dfrac{\partial\bP\bG\bP^T}{\partial\bbeta^T}
                      = \bP \begin{pmatrix}\dfrac{\partial\bG}{\partial\bbeta^T}\end{pmatrix}\bP^T
                      = \bP \Bigl( \text{blkdiag}\big(\bI_{N-1}\otimes\dot\bG^{\beta}_1, \bI_{N-2}\otimes\dot\bG^{\beta}_2, \hdots, \bI\otimes\dot\bG^{\beta}_{N-1} \big)\Bigr) \bP^T
\end{eqnarray}
\setcounter{equation}{\value{eqnCounter4}}
\hrulefill
\end{figure*}
\addtocounter{equation}{2}

The transformation derivatives $\dot{\bG}^{\balpha} \in \mathbb{R}^{\bar{N}L \times \bar{N}L}$, $\dot{\bG}^{\bbeta} \in \mathbb{R}^{\bar{N}L \times \bar{N}L}$ are (\ref{eq:rangeGammaTranslationDerivativeAlpha}) and (\ref{eq:rangeGammaTranslationDerivativeBeta}) respectively, where $\forall\ 1 \le i \le N$, $\dot\bG^{\alpha}_i \in \mathbb{R}^{L \times L}$ and  $\dot\bG^{\beta}_i \in \mathbb{R}^{L \times L}$ are
\begin{align*}
\{ \dot\bG^{\alpha}_i \}_{l+1, \bar{l}+1}=&\  c
\begin{pmatrix} \bar{l} \\ l \end{pmatrix}(-\bar{l})\alpha_i^{-\bar{l}-1}  (-\beta_i)^{\bar{l}-l} \\
\{ \dot\bG^{\beta}_i \}_{l+1, \bar{l}+1} =&\
  \begin{cases}
    \begin{pmatrix} \bar{l} \\ l \end{pmatrix} (\bar{l}-l) \alpha_i^{-l}  (-\beta_i)^{\bar{l}-l-1} & \text{if $\bbeta_i\ne0$} \\
    \bzero_{L, L} & \text{if $\bbeta_i=0$}
  \end{cases}
\end{align*}

\section{Constrained \Cramer\ Rao Bound on distance} \label{ap:ccrbDistance}
The Fisher matrix of $\btheta$ (\ref{eq:FIM}) is \begin{eqnarray}
\bF =\left[
        \begin{array}{c|c|c}
          \bT^T\bT &        \bT^T\bH        & \bT^T\bar{\bV} \\
          \bH^T\bT &        \bH^T\bH        & \bH^T\bar{\bV} \\
          \hline \\[-1em]
          \bar{\bV}^T\bT &  \bar{\bV}^T\bH  & \bar{\bV}^T\bar{\bV}
        \end{array} \right] = \left[
        \begin{array}{c|c}
          \bF_{11}  & \bF^T_{12}  \\
          \hline \\[-1em]
          \bF_{12}  & \bF_{22} \\
        \end{array} \right] \nonumber
\end{eqnarray} and since all the $3$ constraints (discussed in Section \ref{sec:constraintChoice}) are levied on the clock parameters, the orthonormal basis for the null space of these constraints are of the form \begin{eqnarray}
\bU     &=& \left[ \begin{array}{cc} \tilde{\bU}  &  \\  & \bI_{\bar{N}} \end{array} \right].
\end{eqnarray} Following immediately, the CCRB on $\btheta$ is \begin{eqnarray}
\bSigma_{\theta}
&=& \sigma^2\bU\left[ \bU^T\bF\bU\right]^{-1}\bU^T \nonumber \\
&=& \sigma^2\bU\left[ \begin{array}{cc}
                      \tilde{\bU}^T\bF_{11}\tilde{\bU} & \tilde{\bU}^T\bF^T_{12} \\
                      \bF_{12}\tilde{\bU}              & \bF_{22}
                    \end{array} \right]^{-1} \bU^T \nonumber \\
&=& \sigma^2\left[ \begin{array}{cc}
                           *  & * \\
                           * & \cS^{-1}_2
                          \end{array} \right]
=\left[ \begin{array}{cc}
                           *  & * \\
                           * & \bSigma_{\gamma}
                          \end{array} \right]
\end{eqnarray} where $\bSigma_{\gamma}$ is the lower bound on $\bgamma$ and the Schur complement $\cS_2$ is given by \begin{eqnarray}
\cS_2  &=& \bF_{22} -
            \bF_{12}\tilde{\bU}\left[\tilde{\bU}^T\bF_{11}\tilde{\bU}\right]^{-1} \tilde{\bU}^T\bF^T_{12} \nonumber \\
       &=& \bar{\bV}^T\bar{\bV} -
            \bF_{12}\tilde{\bU}\left[\tilde{\bU}^T\bF_{11}\tilde{\bU}\right]^{-1} \tilde{\bU}^T\bF^T_{12},
\end{eqnarray} and subsequently, the CCRB on distance $\bd$ is given by \begin{equation}
\bSigma_d=  c^2\bV\bSigma_{\gamma}\bV=  c^2\sigma^2\bV\cS^{-1}_2\bV^T.
\end{equation} It is observed that contribution of the term $\bF_{12}\tilde{\bU}\left[\tilde{\bU}^T\bF_{11}\tilde{\bU}\right]^{-1} \tilde{\bU}^T\bF^T_{12}$ is insignificant (in all 3 constraint cases) for all practical values of clock, distance and time measurement values under classical two-way time stamp exchange assumption. Hence, the CCRB of distance and the performance of the MGLS solution is observed to be independent of the clock constraints in the simulations.

\newpage
\bibliographystyle{core/IEEEtran}
\bibliography{IEEEabrv,core/myRef, core/olfar}
\end{document}